\documentclass[a4paper,fleqn,usenatbib]{mnras}

\usepackage{newtxtext,newtxmath}

\usepackage[T1]{fontenc}
\usepackage{ae,aecompl}

\usepackage{graphicx} 
\usepackage{float}
\usepackage{amsmath}	
\usepackage{amssymb}	
\usepackage{dcolumn}
\usepackage{epsfig}
\usepackage{color}
\usepackage{pdflscape}
\usepackage{upgreek}
\usepackage{mathptmx}
\usepackage{grffile}
\usepackage{array,float}
\usepackage{multirow}
\usepackage{lscape}
\usepackage{hyperref}
\usepackage{xargs}
\usepackage{xcolor}
\usepackage[normalem]{ulem}
\usepackage{tikz}
\usetikzlibrary{shapes,arrows}
\usepackage{threeparttable}
\usepackage{verbatim}
\usepackage{longtable}

\usepackage{soul,xcolor}

\newcolumntype{d}[1]{D{.}{\cdot}{#1}}
\newcolumntype{.}{D{.}{.}{-1}}

\newcommand{\lsun}{L$_\odot$}
\newcommand{\msun}{M$_\odot$}

\newcommand{\lmratio}{\emph{L}$_{\rm{bol}}$/\emph{M}$_{\rm{fwhm}}$}

\newcommand{\kms}{km\,s$^{-1}$}
\newcommand{\cmthree}{cm$^{-3}$}
\newcommand{\cmtwo}{cm$^{-2}$}

\newcommand{\hi}{H{\sc i}}
\newcommand{\hii}{H{\sc ii}}

\newcommand{\htwoo}{H$_2$O}
\newcommand{\nhthree}{NH$_3$~}

\tikzstyle{decision} = [diamond, draw, fill=blue!20, text width=4.5em, text badly centered, node distance=3cm, inner sep=0pt, text=black, draw=black]
\tikzstyle{block} = [rectangle, draw, fill=blue!20, text width=5em, text centered, minimum height=4em, text=black, draw=black]
\tikzstyle{line} = [draw, -latex', text=black, draw=black, fill=black]
\def\firstellip{(1.6, 0) ellipse [x radius=3cm, y radius=1.5cm, rotate=50]}
\def\secondellip{(0.3, 1cm) ellipse [x radius=3cm, y radius=1.5cm, rotate=50]} 
\def\thirdellip{(-1.6, 0) ellipse [x radius=3cm, y radius=1.5cm, rotate=-50]} 
\def\fourthellip{(-0.3, 1cm) ellipse [x radius=3cm, y radius=1.5cm, rotate=-50]} 

\title[Interstellar masers]{ATLASGAL - Relationship between dense star forming clumps and interstellar masers}

\author[S.\,J.\,Billington et al.]{
	S.\,J.\,Billington,$^{1}$\thanks{E-mail: sjbb\_astro@mail.com} J.\,S.\,Urquhart,$^{1}$
	C.\,K\"{o}nig,$^{2}$ H.\,Beuther,$^{3}$ S.\,L.\,Breen,$^{4}$
	K.\,M.\,Menten,$^{2}$  \newauthor J.\,Campbell-White,$^{5}$ S.\,P.\,Ellingsen,$^{6}$ M.\,A.\,Thompson,$^{7}$ T.\,J.\,T.\,Moore,$^{8}$ D.\,J.\,Eden,$^{8}$ \newauthor W.\,-J.\,Kim$^{9}$ and S.\,Leurini$^{2,10}$ \\
	\\
	$^{1}$ Centre for Astrophysics and Planetary Science, University of Kent, Canterbury CT2\,7NH, UK \\
	$^{2}$ Max-Planck-Institut f\"{u}r Radioastronomie, Auf dem Higel 69, D-53121 Bonn, Germany \\
	$^{3}$ Max Planck Institute for Astronomy, K\"{o}nigstuhl 17, 69117 Heidelberg, Germany \\
	$^{4}$ SKA Organisation, Jodrell Bank Observatory, SK11 9DL, UK\\
	$^{5}$ SUPA, School of Science and Engineering, University of Dundee, Nethergate, Dundee DD1 4HN, U.K.\\
	$^{6}$ School of Mathematics and Physics, University of Tasmania, Private Bag 37, Hobart, Tasmania 7001, Australia \\
    $^{7}$ Science and Technology Research Institute, University of Hertfordshire, College Lane, Hatfield, AL10 9AB, UK \\
	$^{8}$ Astrophysics Research Institute, Liverpool John Moores University, Liverpool Science Park, 146 Brownlow Hill, Liverpool, L3\,5RF, UK \\
	$^{9}$ Instituto de Radioastronom\'{i}a Milim\'{e}trica (IRAM), Granda, Spain \\
	$^{10}$ INAF-Osservatorio Astronomico di Cagliari, Via della Scienza 5, I-09047, Selargius (CA)
}
\date{Accepted XXX. Received YYY; in original form ZZZ}

\pubyear{2019}

\begin{document}
\label{firstpage}
\pagerange{\pageref{firstpage}--\pageref{lastpage}}
\maketitle

\begin{abstract}

\vspace{-0.1cm}

We have used catalogues from several Galactic plane surveys and dedicated observations to investigate the relationship between various maser species and Galactic star forming clumps, as identified by the ATLASGAL survey. The maser transitions of interest are the 6.7 \& 12.2\,GHz methanol masers, 22.2\,GHz water masers, and the masers emitting in the four ground-state hyperfine structure transitions of hydroxyl. We find clump association rates for the water, hydroxyl and methanol masers to be 56, 39 and 82\,per\,cent respectively, within the Galactic longitude range of $60\degr > \ell > -60\degr$. We investigate the differences in physical parameters between maser associated clumps and the full ATLASGAL sample, and find that clumps coincident with maser emission are more compact with increased densities and luminosities. However, we find the physical conditions within the clumps are similar for the different maser species. A volume density threshold of $n$(H$_2$) $>$ 10$^{4.1}$\,\cmthree\ for the 6.7\,GHz methanol maser found in our previous study is shown to be consistent across for all maser species investigated. We find limits that are required for the production of maser emission to be 500\,\lsun\ and 6\,\msun\ respectively. The evolutionary phase of maser associated clumps is investigated using the $L$/$M$ ratio of clumps coincident with maser emission, and these have similar $L$/$M$ ranges ($\sim10^{0.2} - 10^{2.7}$ \lsun/\msun) regardless of the associated transitions. This implies that the conditions required for the production of maser emission only occur during a relatively narrow period during a star's evolution. Lower limits of the statistical lifetimes for each maser species are derived, ranging from $\sim$ 0.4 $-$ 2$\times$10$^{4}$\,yrs and are in good agreement with the ``straw man'' evolutionary model previously presented.

\end{abstract}



\section{Introduction}
\label{sect:intro}

Masers exist across the Galaxy and the different species of masers have been shown to trace certain physical processes and are associated with particular celestial objects (e.g. late-type stars, \hii\ regions, star-forming regions; \citealt{Elitzur1992}). Over the last decade, research has been devoted to the study of maser emission, where large-scale surveys have resulted in the production of large, comprehensive and statistically representative catalogues of the various maser species (e.g. \citealt{Caswell2010, Walsh2014, Qiao2016, Qiao2018, Beuther2019}). Catalogues of maser species have driven detailed investigations and theoretical studies into the environments and conditions required for their production \citep{Sobolev1997, Norris1998, Cragg2001, Breen2010, Urquhart2013}. Masers have been shown to be good chemical probes of star forming environments \citep{Menten1997} and different species have been regularly shown to be coincident with each other in positional space \citep{Forster1989,Menten1992,Menten1997}.

The class II 6.7\,GHz and 12.2\,GHz methanol (CH$_3$OH) masers, first reported by \cite{Menten1991} and \cite{Batrla1987} respectively, have been found to be exclusively associated with intermediate- and high-mass young stellar objects (YSO; \citealt{Minier2003,Xu2008, Breen2013, Urquhart2015, Billington2019a}). Therefore, the 6.7 and 12.2\,GHz methanol masers provide a simple and convenient method for identifying potential regions of massive star formation. A number of observations have been made towards these maser species and in-depth studies have been conducted (e.g. \citealt{Menten1992,Phillips1998,Szymczak2002,Blaszkiewicz2004,Goedhart2005,Green2009}). It is thought that the class II methanol masers are produced in the accretion disks around young massive stellar objects \citep{Norris1998}, and are radiatively pumped in small, confined regions with temperatures of $\sim$150\,K, methanol column densities of $>$10$^{15}$\,\cmtwo\ and volume densities of $>$10$^8$\,\cmthree\ \citep{Sobolev1997,Cragg2001}. However, other studies have proposed different scenarios, such as the existence of these maser species within expanding shock waves or protostellar outflows \citep{Walsh1998}. Although, it could in fact be true that methanol masers exist in all of these situations \citep{Minier2000}.

There also exist other maser species that are known to be associated with regions of star formation, such as the 22.2\,GHz maser emission produced by water (H$_2$O; \citealt{Cheung1969}), and the four ground-state transitions of hydroxyl (OH; \citealt{Weinreb1963}) at 1612, 1665, 1667 and 1720\,MHz. Observations of water masers indicate that these are found in outflows as well as circumstellar disks of both low- and high-mass young stellar objects \citep{Claussen1996, Codella2004, Titmarsh2014, Titmarsh2016}, and are associated with regions of ongoing star formation. \cite{Elitzur1989} presented a model for H$_2$O masers in star forming regions, where masers are produced in post-shocked gas with initial densities of $n \sim$10$^{7}$\,\cmthree\ and temperatures of $\sim$400\,K. Hydroxyl masers are known to trace a number of different environments. The most abundant transition at 1612\,MHz is typically found in the expanding shells of evolved stars \citep{Wilson1968, Elitzur1976}, while the emission lines of 1665 and 1667\,MHz have generally been found towards regions of star formation \citep{Argon2000, Qiao2014, Qiao2016}. The 1720\,MHz is the least frequently found and is generally observed towards shock excited regions within star formation sites and supernovae remnants \citep{Claussen1999, Caswell2004}. However, none of these associations are exclusive and all four transitions can be found within different environments and even toward the same region \citep{Caswell2013, Walsh2016, Beuther2019}. \cite{Cragg2002a} produced models of the OH masers, and found that OH masers require temperatures of $\sim$100\,K, moderately high densities of $n$ = 10$^5$ $-$ 10$^8$\,\cmthree, conditions which are likely to be met in all regions where OH masers are found.

As these three maser species are capable of tracing different physical conditions and processes, studies have been undertaken to investigate the relationship between the masers and environments where high-mass stars are born (e.g. \citealt{Beuther2002, Breen2010, Breen2011, Urquhart2013, Urquhart2015}). Studies have also investigated the relationship between different maser species (e.g. \citealt{Forster1989,Menten1997,Szymczak2005,Jones2020}), and it has been shown that different maser transitions are often spatially associated with one another (within 15\,arcsec). \cite{Ellingsen2007b} presented a ``straw man'' model for the evolution and relative timescales of different maser species (water, methanol, hydroxyl) within regions of star formation. This model was further refined by \cite{Breen2010} who estimated relative lifetimes for each of the maser species. The sequence of events as described by this model state that during star formation, class I (collisionally pumped) methanol masers occur first, followed by the appearance of water masers, produced in the post-shocked gas due to outflows associated with these object. Class II (radiatively pumped) masers are then produced in accretion disks due to the thermal output of the natal star. Finally, hydroxyl masers are seen towards developing \hii\ regions, and as these \hii\ regions evolve, they disrupt and disperse their environments, destroying the conditions necessary for any type of maser emission to exist (as indicated by the decreased detection rate of all maser types in the late evolutionary phases).

In three previous papers (\citealt{Urquhart2013, Urquhart2015, Billington2019a}), we have presented a detailed analysis of the 6.7\,GHz methanol masers, as identified by the Methanol MultiBeam survey (MMB; \citealt{Green2009a}), and submillimetre dust emission, as identified by the APEX Telescope Large Area Survey of the Galaxy (ATLASGAL; \citealt{Schuller2009}). These works identified the almost ubiquitous association between the 6.7\,GHz maser and dust emission ($\sim$99\,per\,cent), and showed that this maser species has a well defined turn-on and termination point, existing for a distinct amount of time in the evolutionary timescale of massive star formation, with a statistical lifetimes of 3.3$\times$10$^4$\,yrs \citep{Billington2019a}. The findings of this study supports the theoretically derived value (\citealt{VanDerWalt2005}; 2.5 $-$ 4.5$\times10^4$\,yrs). Given the availability of large scale water and hydroxyl surveys, it is possible to extend our previous work and fully test the ``straw man'' model presented by \cite{Ellingsen2007b} and \cite{Breen2010} by investigating the lifetimes and evolutionary timescales of multiple maser species.

This study will extend the work of \cite{Billington2019a} by investigating the physical properties of dense Galactic clumps associated with each maser transition, for which catalogues have been produced from a number of Galactic surveys. The Galactic surveys of interest are: the MMB survey \citep{Caswell2010,Green2010,Caswell2011a,Green2012,Breen2015}, The \hi/OH/Recombination line survey (THOR; \citealt{Beuther2016,Beuther2019}), the Southern Parkes Large-Area Survey in Hydroxyl (SPLASH; \citealt{Dawson2014, Qiao2014, Qiao2016}), the \htwoo\ southern Galactic Plane Survey (HOPS; \citealt{Walsh2011,Walsh2014}) and observations of 12.2\,GHz methanol masers undertaken by \cite{Breen2012,Breen2012b,Breen2014a,Breen2016a}. These surveys are described in Section\,\ref{sect:surveys}.  Along with investigating the physical properties of maser associated clumps, we derive statistical lifetimes for each of the masers species and identify their positions in the evolutionary timescales of star formation.

The structure of this paper is as follows: in Section\,\ref{sect:surveys}, we describe the archival survey data that will be used throughout the study. In Section\,\ref{sect:matching} we describe the matching procedure used for the maser species of interest. Section\,\ref{sect:parameters} presents the derivation of the various physical parameters of maser associated dust clumps, along with the corresponding uncertainties. In Section\,\ref{sect:discussion} we discuss various aspects of the different maser samples, along with presenting the statistical lifetimes for each maser species. A summary of our findings is given in Section\,\ref{sect:conclusions}.

\begin{figure}
	\includegraphics[width=0.5\textwidth]{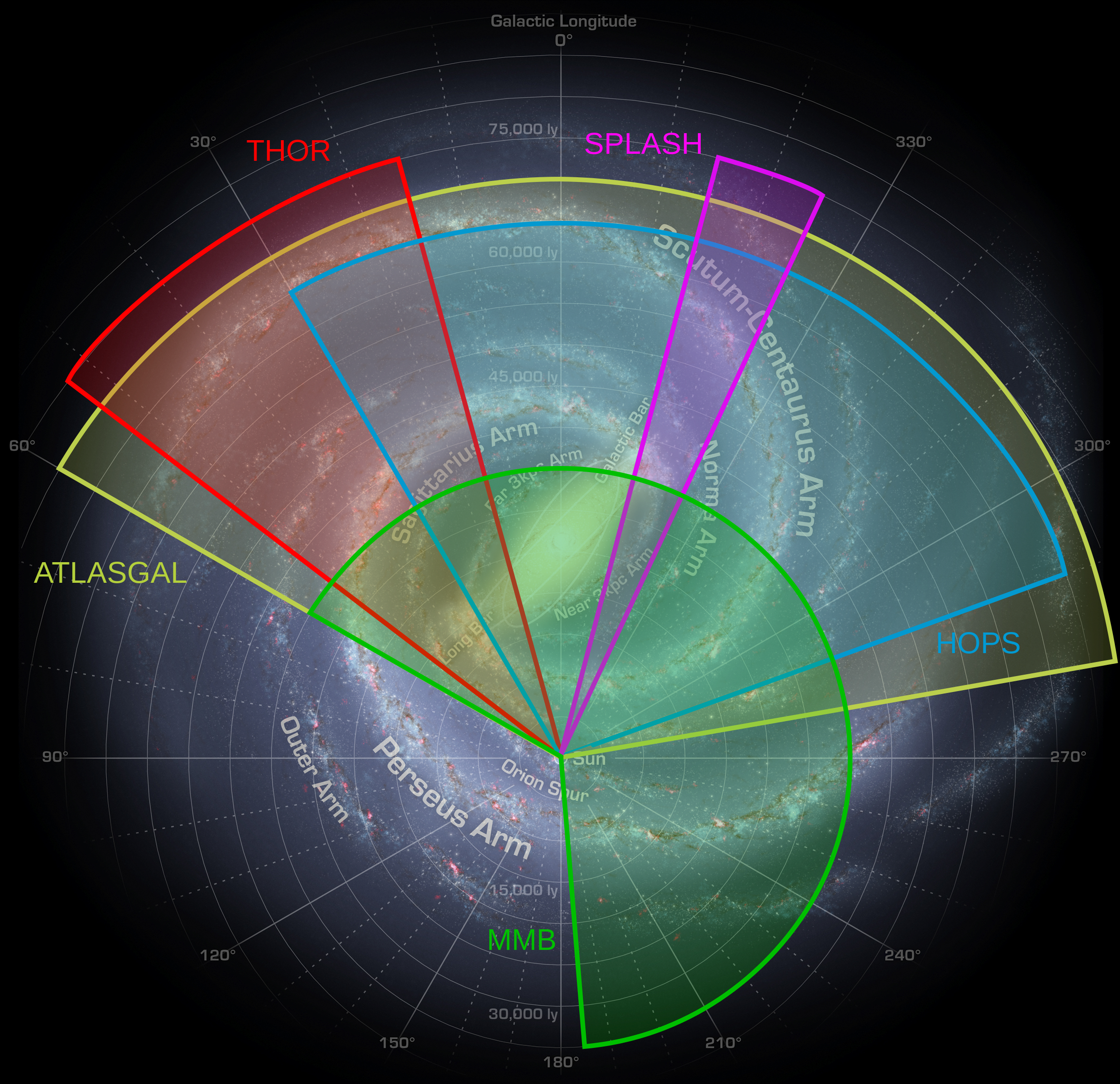}
	\caption{Image depicting the Galactic longitude coverages for the Galactic plane surveys that are utilised in this study (courtesy of NASA/JPL-Caltech/R. Hurt (SSC/Caltech). The approximate coverage of each survey is shown along with each surveys' corresponding name. The length of each sector is arbitrary.}
	\label{fig:survey_coverages}
\end{figure}

\section{Survey Descriptions}
\label{sect:surveys}

Figure\,\ref{fig:survey_coverages} shows the survey coverages for each of the surveys that are utilised in this study. In Table\,\ref{table:survey_coverages} we present a summary of the coverages for each survey.

\subsection{Descriptions of maser surveys}

\subsubsection{The \htwoo\ southern Galactic Plane Survey (HOPS)}

The \htwoo\ southern Galactic Plane Survey (HOPS; \citealt{Walsh2011}) is an unbiased survey of 100 sq. degrees of the Galactic plane ($30\degr \geq \ell \geq 290\degr$, $|b| < 0.5\degr$). HOPS was performed using the Australia Telescope National Facility (ATNF) Mopra 22\,m radio telescope and detected 540 sites of 22.2\,GHz H$_2$O maser emission \citep{Walsh2011}. \cite{Walsh2014} performed follow-up observations of these maser sites using the Australia Telescope Compact Array (ATCA), to provide accurate positions for individual maser spots within 1\,arcsec. The typical noise for these observations was 15$-$167\,mJy. The higher resolution observations identified 631 maser sites (collections of maser spots) in the 540 regions detected by Mopra; these consisted of 2790 individual spectral features (maser spots).

Out of the 631 maser sites detected in \cite{Walsh2014}, 433 were identified as being associated with star formation, 121 sites with evolved stars and the remaining 77 were classified as unknown. They also found that evolved stars tend to have more maser spots than those associated with star formation. Furthermore, they found that maser sites associated with evolved stars are much more clustered around localised positions than the full area of star formation regions. The linear positions of maser spots within star formation regions present features that are both parallel and perpendicular to known outflows, where H$_2$O masers are thought to originate.

\subsubsection{The \hi/OH/Recombination (THOR) line survey}

The \hi/OH/Recombination (THOR; \citealt{Beuther2016}) line survey of the inner Milky Way has surveyed $14.5\degr \leq \ell \leq 66.8\degr$ and $|b| < 1.25\degr$ of the Galactic plane using the Karl Jansky Very Large Array (VLA) in the C-array configuration. The THOR survey has observed a number of different spectral lines, including the \hi\ 21\,cm, nineteen of the H$n\alpha$ radio recombination lines and the four ground-state transitions of the hydroxyl maser, as well as continuum emission between 1 and 2\,GHz \citep{Bihr2016, Wang2018}. The survey has angular resolutions of $\sim$ 12.5 to 19\,arcsec, with a spectral resolution of 1.5\,\kms, and typical rms noise levels of $\sim$10\,mJy.

Recently, THOR has produced an unbiased catalogue of the four OH maser emission lines at 1612, 1665, 1667 and 1720\,MHz (\citealt{Beuther2019}). The catalogue is comprised of OH maser sources in the northern hemisphere, tracing a number of different astronomical phenomena, as previously described in Sect.\,\ref{sect:intro}. We do note here that within the THOR survey, the 1667\,MHz maser was not covered in $\sim$50\,per\,cent of the survey (below $\ell$ = 29.2$\degr$, between 31.5$\degr$ and 37.9$\degr$ and between 47.1$\degr$ to 51.2$\degr$), however, this will not affect the results and statistics presented in this study.

\subsubsection{The Southern Parkes Large-Area Survey in Hydroxyl (SPLASH)}

The Southern Parkes Large-Area Survey in Hydroxyl (SPLASH) is an unbiased southern Galactic plane survey of the four ground-states transitions of the hydroxyl molecule, undertaken using the Parkes 64-m telescope and utilising the H-OH receiver. The first results of this survey are presented in \cite{Dawson2014} and covers the Galactic plane between longitudes 334\degr\ and 344\degr\ and latitudes $|b|$ < 2\degr. Within this region, the SPLASH survey has detected a total of 495 masers, across the four ground transitions of hydroxyl. 

\cite{Qiao2016} presented accurate positions of the masers identified within the pilot region. These observations were undertaken using the Australia Telescope Compact Array, with angular resolutions of between 4$-$13\,arcsec, spectral resolution of 0.09\,\kms\ and typical rms noise levels of $\sim$65\,mJy. The catalogue presented in \cite{Qiao2016} provides a further sample of hydroxyl masers in the southern Galactic plane and we use these to supplement the THOR survey maser catalogue.

\subsubsection{Class II Methanol masers (6.7 \& 12.2\,GHz)}

The Methanol Multibeam (MMB; \citealt{Green2009a}) survey has surveyed the Galactic plane between $60\degr \geq \ell \geq 186\degr$ and $|b| < 2\degr$, and has produced a catalogue of 972 emission sites of the class II 6.7\,GHz methanol maser across the Galaxy \citep{Caswell2010,Green2010,Caswell2011a,Green2012,Breen2015}. Initial detections were made using the ATNF Parkes 64\,m radio telescope, and were followed up with ATCA or the Multi-Element Radio Linked Interferometer Network (MERLIN; \citealt{Thomasson1986}) to determine accurate positions ($<$1\,arcsec) to maser emission sites if those positions were not available in the literature. The average noise in survey cubes was 170\,mJy. It is likely that the MMB survey accounts for the majority of 6.7\,GHz methanol masers across the Galaxy, as it has been shown that away from the Galactic plane, 6.7\,GHz maser emission is rare \citep{Yang2017,Yang2019}. This is also consistent with them being only associated with high-mass star formation.

\cite{Breen2012,Breen2012b,Breen2014a,Breen2016a} have produced a number of MMB follow-up observations towards sites of 6.7\,GHz methanol masers, searching for the class II 12.2\,GHz methanol maser transition. The 12.2\,GHz maser is the strongest and most widespread class II methanol maser line after the 6.7\,GHz transition, and over the same coverage as the MMB survey ($60\degr \geq \ell \geq 186\degr$), Breen et al. detected 432 12.2\,GHz masers. \citep{Breen2012,Breen2012b,Breen2014a,Breen2016a} also concluded that the 12.2\,GHz maser occurs within star-formation regions at a slightly later stage of evolution than regions that are only associated with the 6.7\,GHz transition. This was explained due to the 12.2\,GHz methanol masers being associated with 6.7\,GHz masers that have relatively higher flux densities and peak luminosities. Furthermore, 12.2\,GHz methanol masers are also more likely to be found towards 6.7\,GHz methanol masers with associated OH masers \citep{Breen2010}. The estimated lifetime of the 12.2\,GHz masers is estimated to be be in the range 1.5$\times 10^4$ and 2.7$\times 10^4$\,yrs \citep{Breen2010}.

It is worth noting that as the 12.2\,GHz maser observations are targeted towards known positions of 6.7\,GHz masers, this sample may not be regarded as an independent sample. However, no previous studies have resulted in the serendipitous detections of 12.2\,GHz methanol masers, and all currently known 12.2\,GHz masers have been found with a 6.7\,GHz counterpart. Also, previous searches \citep{Caswell1995b,Blaszkiewicz2004} have shown that 12.2\,GHz methanol maser emission is rarely brighter than any associated 6.7\,GHz methanol maser emission. Therefore, it is not unreasonable to assume that the observations conducted by \cite{Breen2012,Breen2012b,Breen2014a,Breen2016a} account for the majority of the 12.2\,GHz methanol masers within the MMB survey coverage ($60\degr \geq \ell \geq 186\degr$).

\subsection{ATLASGAL survey}

The Apex Telescope Large Area Survey of the GALaxy (ATLASGAL; \citealt{Schuller2009}) has mapped dust emission at 870\,\micron\ across 420 sq. degrees of the Galactic plane, between Galactic longitudes $60\degr \geq \ell \geq 280\degr$ and latitudes $|b| < 1.5\degr$. ATLASGAL is the largest and most sensitive ground-based submillimetre survey to date and has produced a catalogue of over 10\,000 dense clumps within the Galaxy \citep{Contreras2013,Csengeri2014,Urquhart2014a}. The Large APEX Bolometer Camera (LABOCA; \citealt{Siringo2009}) was used to conduct the survey. The APEX telescope has a dish size of 12\,m allowing for a FWHM resolution of 19.2\,arcsec at this wavelength.

\cite{Urquhart2018} produced a catalogue of physical parameters for a large fraction of the ATLASGAL compact source catalogue ($\sim$8\,000), outside of the Galactic centre region ($5\degr < |\ell| < 60\degr$ and $|b| < 5\degr$). The Galactic centre region was omitted from this study to reduce source confusion and difficulty in determining reliable kinematic distances. This catalogue includes representative samples of all of the earliest stages of massive star formation. Following the procedures first established for a representative sample of $\sim$100 ATLASGAL sources \citep{Konig2017}, emission maps from the HiGAL survey \citep{Molinari2010} were used to fit the spectral energy distributions in order to derive dust temperatures and bolometric luminosities of embedded objects. Follow-up molecular line observations of the ATLASGAL emission maps (e.g. CO, NH$_3$, N$_2$H$^+$; see Table\,1 of \citealt{Urquhart2018} for a complete list of observations) have determined radial velocities which, along with the Galactic rotation curve, have been used to derive distances to $\sim$8\,000 regions ($\sim$79\,per\,cent of the full ATLASGAL sample). These distances have been used to determine the properties of dense clumps across the Galaxy as presented in \cite{Urquhart2018}. Every ATLASGAL source has been classified into one of four evolutionary groups based on infrared or radio counterparts and are defined as quiescent, protostellar, young stellar object (YSO) and \hii\ region \citep{Konig2017}.

\begin{table}
	\centering
	\caption{\label{table:survey_coverages}Galactic Longitude coverages for each survey presented in this study.}
	\begin{tabular}{lcccc}
		\hline
		\hline
		Survey Name & Maser Species & Sensitivity (mJy) & Survey Coverage  \\
		\hline
		HOPS        & H$_2$O & 15$-$167 & $30\degr \geq \ell \geq 290\degr$ \\
		THOR        & OH & $\sim$10 & $66.8\degr \geq \ell \geq 14.5\degr$ \\
		SPLASH      & OH & $\sim$65 & $344\degr \geq \ell \geq 334\degr$  \\
		MMB         & Class II CH$_3$OH & $\sim$170 & $60\degr \geq \ell \geq 186\degr$ \\
		ATLASGAL    & 870\,\micron\ continuum & $\sim$60 & $60\degr \geq \ell \geq 280\degr$ \\
		\hline
	\end{tabular}
\end{table}

\begin{figure}
	\includegraphics[width=0.49\textwidth]{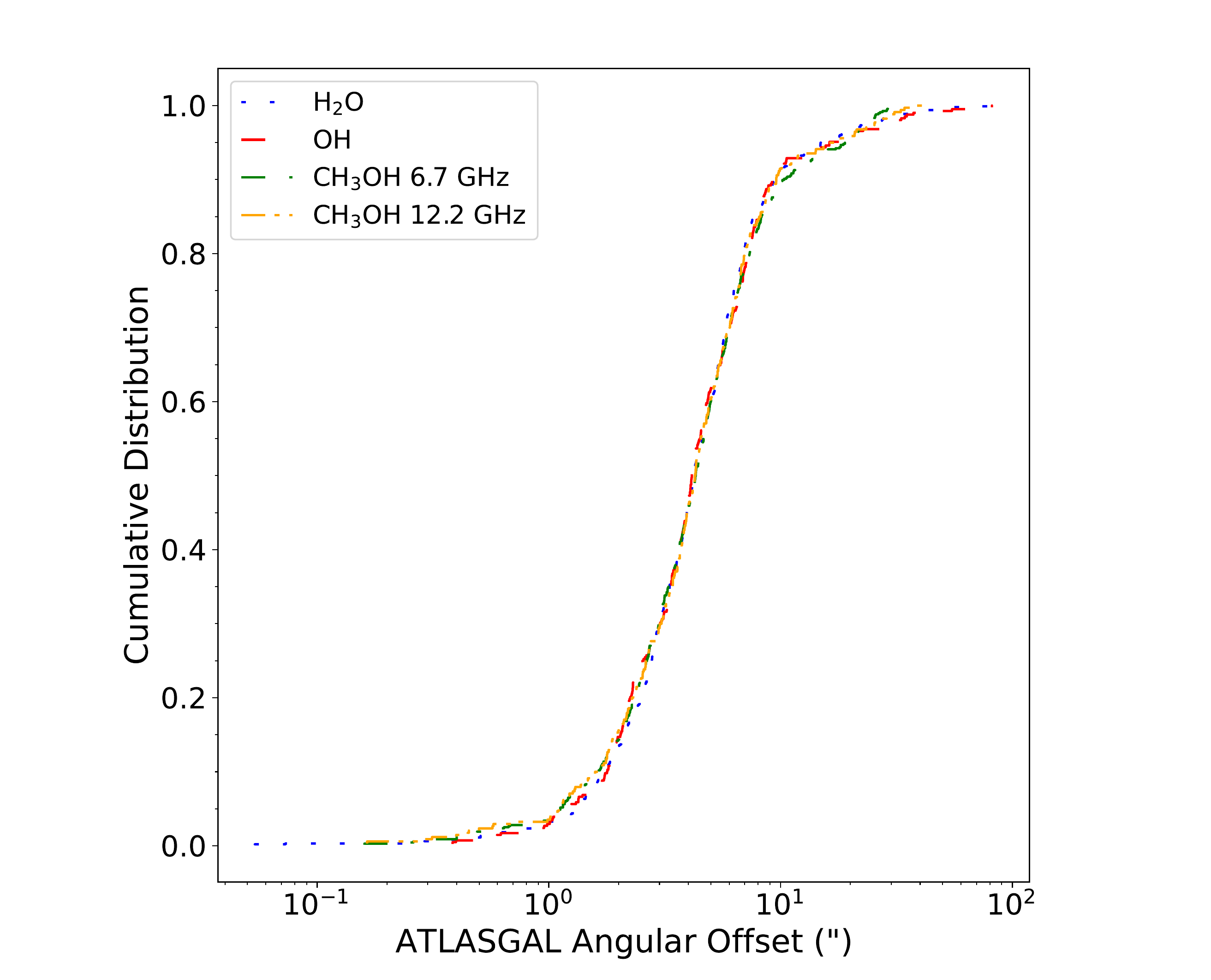}  \\
	\includegraphics[width=0.49\textwidth]{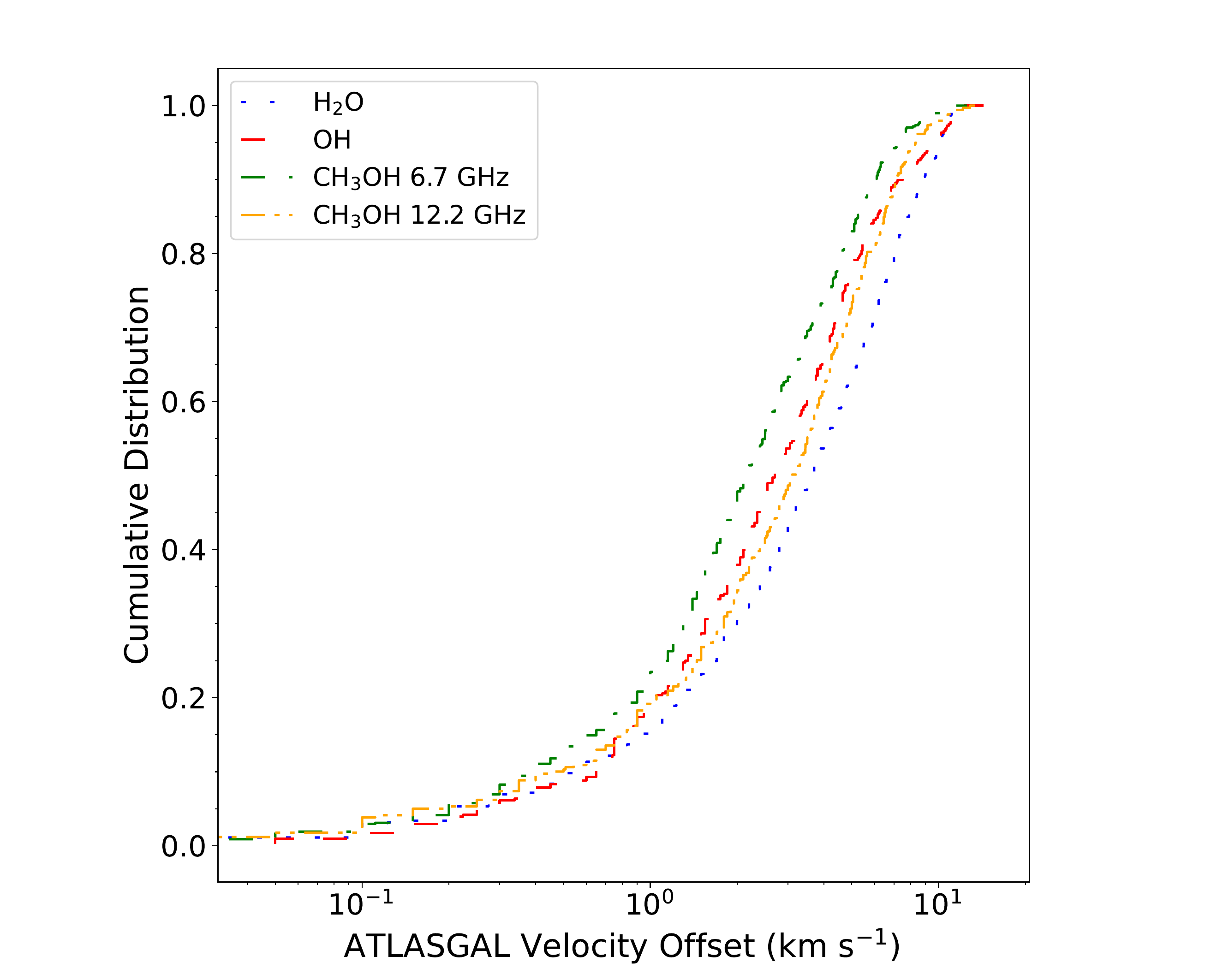}
	\caption{Cumulative distribution function of the maser-ATLASGAL angular and velocity offsets in the upper and lower panels respectively. Each coloured line represents a different maser species as shown by the legend in the top left of each plot. Both panels present the same maser associated clump sample, i.e. those that are within the spatial and velocity matching criteria as explained in the text.}
	\label{fig:offset_cdfs}
\end{figure}

\section{Maser Associations}
\label{sect:matching}

This study will only focus on the Galactic coverage between $5\degr < |\ell| < 60\degr$ and $|b| < 1.5\degr$, the same coverage as \cite{Urquhart2018}, to ensure we have accurate determinations for the physical properties of clumps associated with any maser transition.

By using the catalogues of the surveys mentioned in the previous section, we have attempted to associate multiple maser transitions to dense clumps as found in the ATLASGAL survey. These maser transitions are the 6.7\,GHz \& 12.2\,GHz methanol maser, 22.2\,GHz water maser and the four hydroxyl maser emission lines at 1612, 1665, 1667 and 1720\,MHz. We have utilised the 6.7\,GHz methanol maser matches with the ATLASGAL survey reported in \cite{Urquhart2013} and \cite{Billington2019a} but have not included clumps identified from the JCMT Plane Survey (JPS; \citealt{Moore2015,Eden2017,Billington2019a}) as the data from JPS will provide no further information for this study. We have also chosen not to utilise the targetted observations towards sites of 6.7\,GHz methanol maser emission presented in \cite{Urquhart2015} that were beyond the boundaries of the ATLASGAL survey. 

When referring to a specific maser species or transition in this study, the corresponding frequency of that maser will always be given, in contrast, the term "maser" used independently will refer to all the maser transitions.

\subsection{Spatial Matching}

Initial matches were identified using a 90\,arcsec\ search radius between maser emission and the 870\,\micron\ peak dust emission. This radius is the 3$\sigma$ value of the effective radius distribution for ATLASGAL sources and so 99.7\,per\,cent of dust continuum sources have a radius of less than this size. Images have been created from the ATLASGAL 870\,\micron\ emission maps, so that each match could be confirmed visually. A match was confirmed if the maser source was within the 3$\sigma$ boundary of the 870\,\micron\ emission and all masers found beyond this boundary have been removed from any further analysis, as these are unlikely to be associated with star formation. We find 549 OH (429 from THOR and 120 from SPLASH), 359 water and 392 12.2\,GHz masers sites that are spatially coincident with dust emission (within 3$\sigma$). A cumulative distribution function of the angular offsets for each of the maser species can be found in the upper panel of Fig.\,\ref{fig:offset_cdfs}. It can be seen from this Figure that the majority of masers ($\sim$90\,per\,cent) lie within 10\,arcsec of a dust continuum peak. As the majority of star formation is concentrated towards the highest column density regions in the centre of clumps \citep{Urquhart2014b}, this provides further confirmation that the masers are associated with star formation. It can also be seen from the upper panel in Fig\,\ref{fig:offset_cdfs} that there is no significant differences in the spatial offsets between the various maser species. However, they may be differences on scales less than 10\,arcsec but the resolution of the data is not sufficient to investigate this.

\subsection{Velocity Offsets}

Molecular line velocities of the dense clumps are available for the majority of masers and for a large portion of the ATLASGAL survey (see Table\,1 of \citealt{Urquhart2018} for a complete list of observations and references). We use these measurements to examine the correlation between the masers and the dense clumps to confirm their associations. The difference between maser median velocity and clump median velocity have been calculated for all of the maser species that are spatially coincident. The cumulative distribution functions of these distributions can be found in the lower panel of Fig.\,\ref{fig:offset_cdfs}. It can be seen from this Figure that water masers appear to have larger velocity offsets, which is consistent with this type of maser being associated with protostellar outflows.

\begin{figure}
	\includegraphics[width=0.49\textwidth]{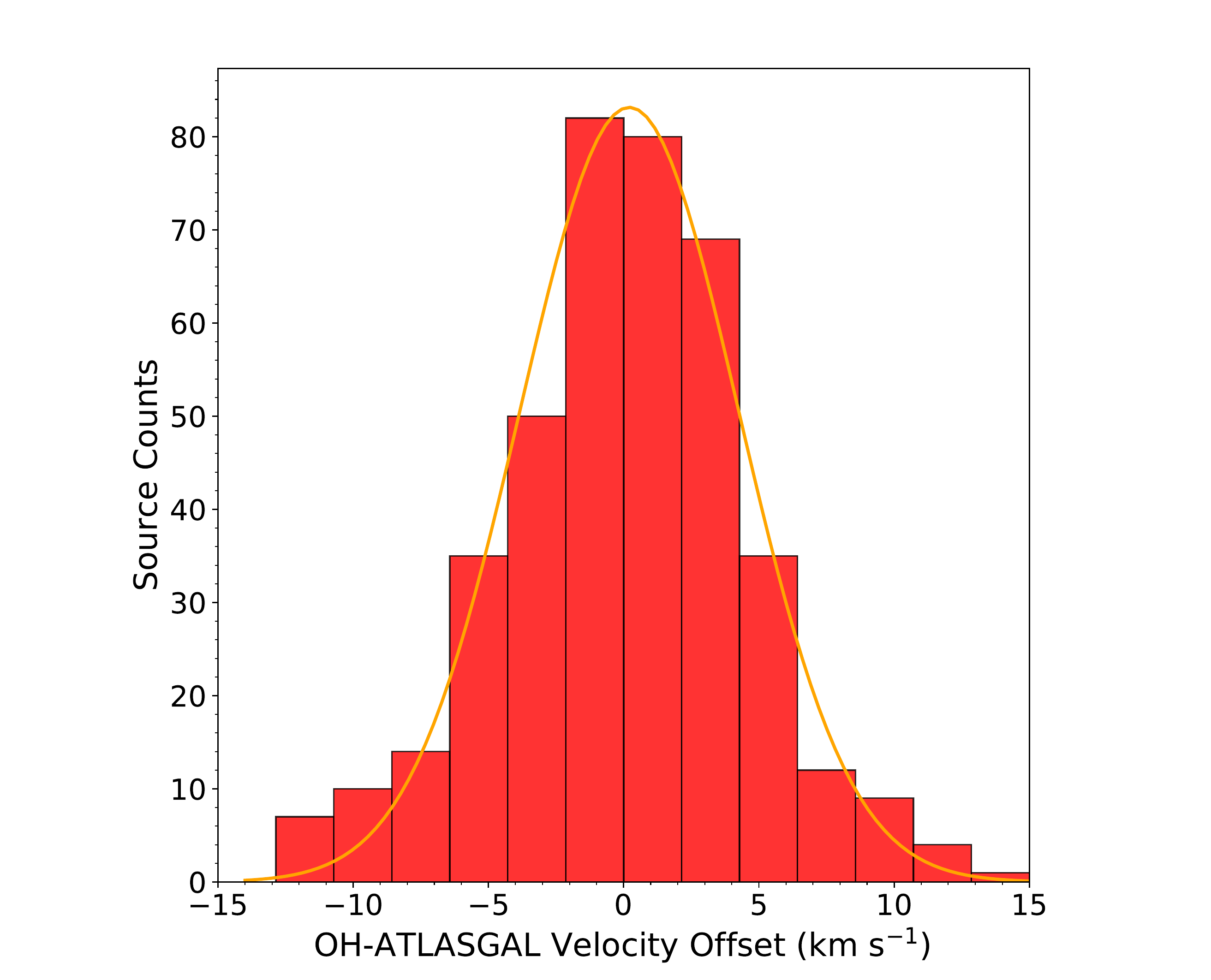}
	\includegraphics[width=0.49\textwidth]{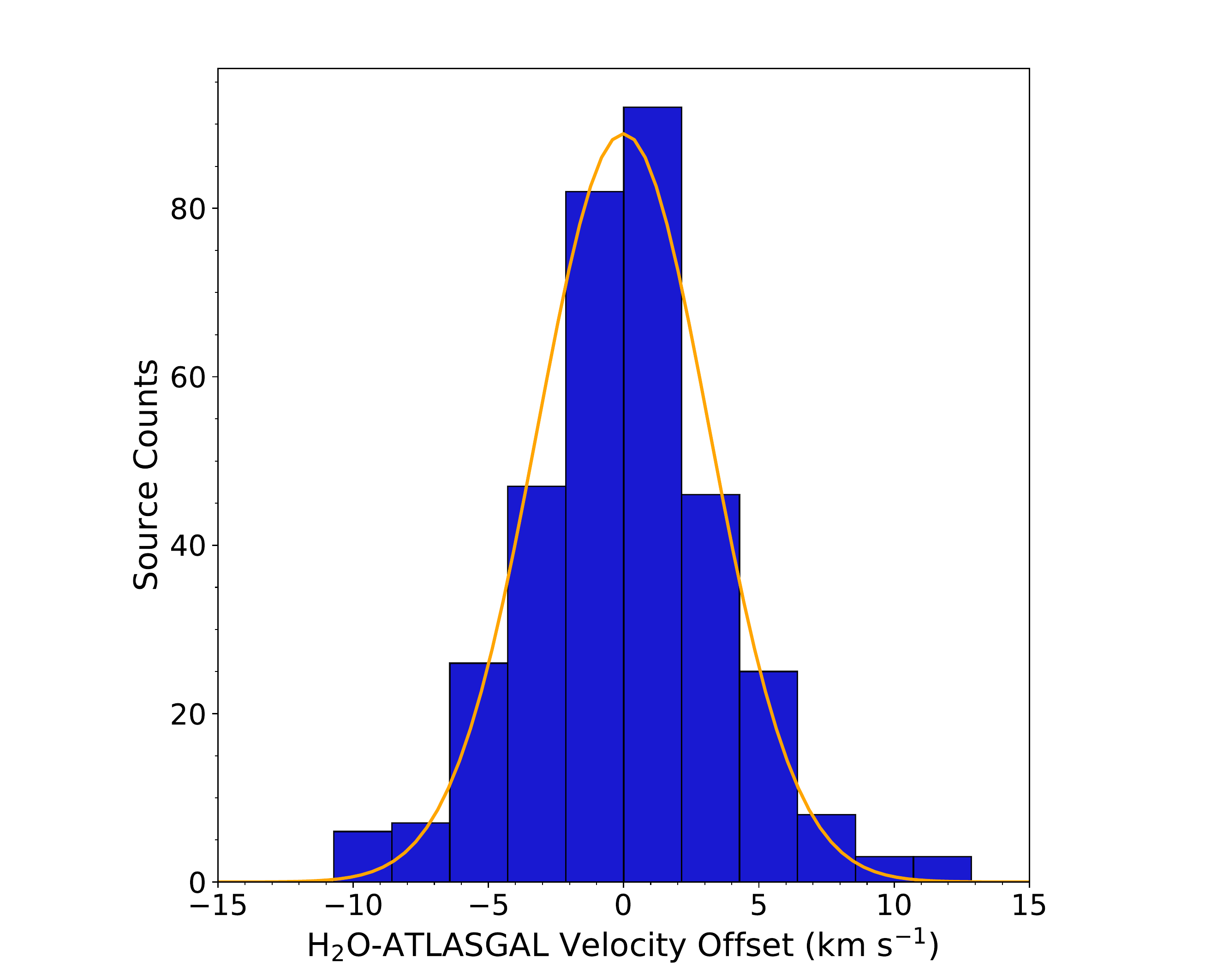} \\
	\includegraphics[width=0.49\textwidth]{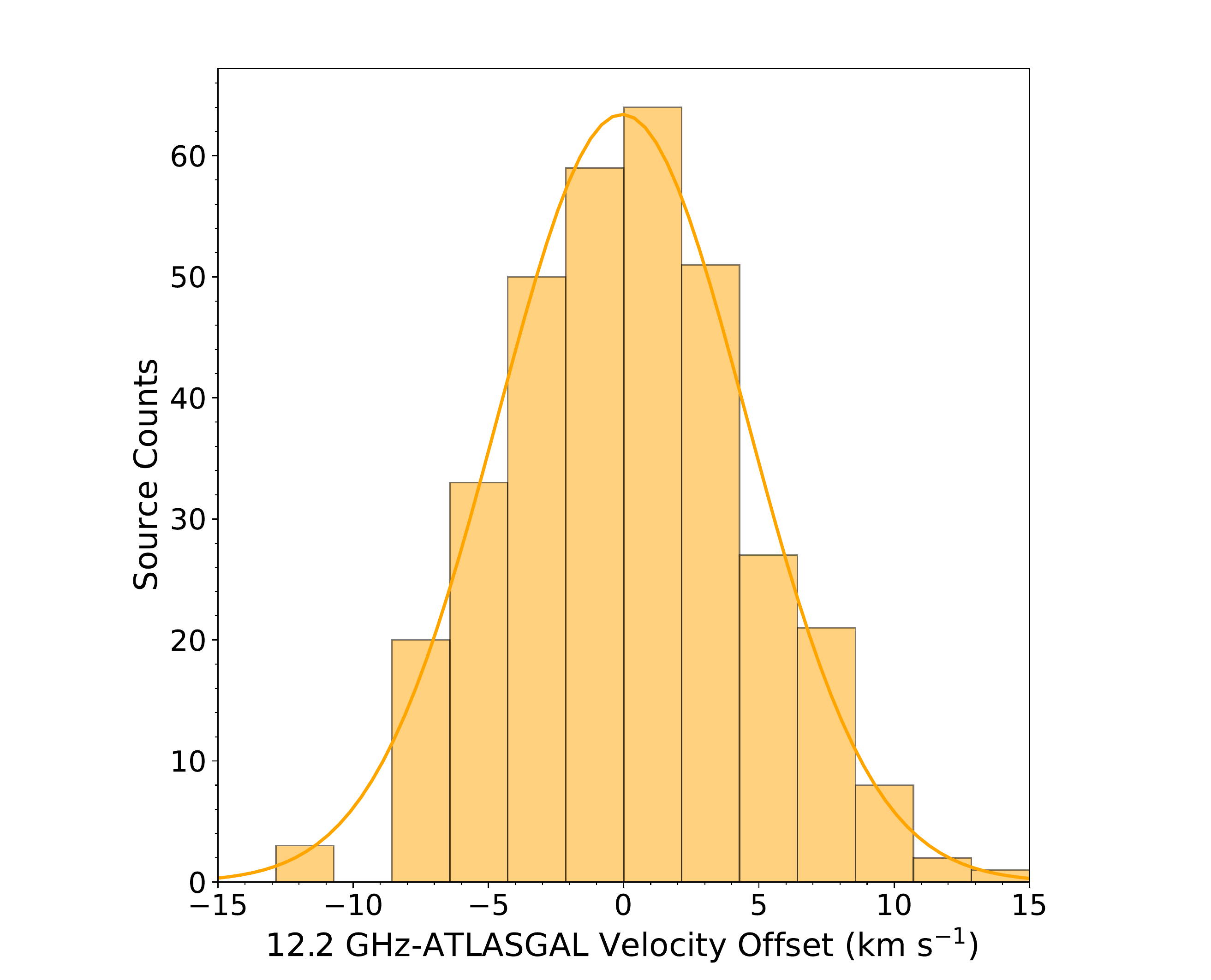} \\
	\caption{Histograms presenting the velocity offsets between the median maser velocities and the ATLASGAL molecular line velocities. The panels present the hydroxyl, water and 12.2\,GHz methanol maser velocity distributions in the upper, middle and lower panels, respectively. The distributions have been truncated at $-$15 and 15\,\kms. A Gaussian profile has been fitted to each histogram and is shown as the solid yellow lines in each panel.}
	\label{fig:velo_offsets}
\end{figure}

In Fig.\,\ref{fig:velo_offsets} we show the velocity offsets for the different maser species and these have been fitted with Gaussian profiles. The Gaussian fits appear to be a reasonable model for all of the velocity distributions. To confirm our maser associated sample, we have opted to only include clumps with a maser source that has a median velocity difference of less than 3$\sigma$, which is calculated from the Gaussian models. This likely accounts for any possible line-of-sight alignments between masers and clumps that may not be spatially coincident. It is likely that masers with a high velocity difference compared to a spatial aligned clump, may in fact not be associated with each other. Since the focus of this study is on the dense clumps associated with maser emission and not the masers themselves, this will help to refine the true sample of maser associated clumps. The 3$\sigma$ values for the different distributions are 11.56, 11.92 and 13.89\,\kms\ for the \htwoo, OH and 12.2\,GHz methanol masers respectively. The 3$\sigma$ value for the 6.7\,GHz methanol masers is taken from \cite{Billington2019a} as 13.5\,\kms.

\subsection{Total matched sample}

\begin{table}
	\caption{\label{table:maser_matches}Total matches for each maser species and the ATLASGAL catalogue between Galactic longitudes $60\degr > \ell > -60\degr$, where all matches are below our 3$\sigma$ threshold in both positional and velocity space. The percentage of associated dust clumps within each survey coverage is also given. Errors have been calculated using Binomial statistics.}
	\begin{tabular}{lcc}
		\hline
		\hline
		Maser Transition & Maser-Dust              & Percentage of            \\ 
		                 & Associations            & associated dust clumps$^{1}$   \\ 
		\hline
		CH$_3$OH 6.7\,GHz  & 839/918 (91 $\pm$ 0.9\%)   & 8.5\%    \\
		CH$_3$OH 12.2\,GHz & 340/414 (82 $\pm$ 1.9\%)   & 3.2\%    \\
		H$_2$O 22.2\,GHz   & 345/614 (56 $\pm$ 1.4\%)   & 2.0\%    \\
		OH 1612\,MHz       & 43/1378 (3 $\pm$ 0.5\%)   & 0.9\%     \\
		OH 1665\,MHz       & 238/394 (60 $\pm$ 2.5\%)   & 3.6\%    \\
		OH 1667\,MHz       & 96/226  (42 $\pm$ 3.2\%)  & 1.2\%     \\
		OH 1720\,MHz       & 31/63   (49 $\pm$ 6.3\%)  & 0.7\%     \\
		\hline
	\end{tabular}
	$^{1}$ Number of ATLASGAL sources with an associated detected maser
\end{table}

We have matched a number of different maser species to dense clumps across the Galaxy as identified by the ATLASGAL survey. Our sample has been constrained using the positional and velocity offsets as previously described. The number of maser associations for each maser transition can be found in Table\,\ref{table:maser_matches}.

We find that there is no difference in the spatial offsets between each maser species and their associated dust clump. The average offset between any maser species and the peak dust emission is $\sim$\,6\,arcsec with the 90\,per\,cent range of the maser samples having offsets between 1.7\,arcsec and 10.3\,arcsec. The absolute velocity differences for all of the maser species range between 2.8$-$4.3\,\kms\ with the 6.7\,GHz methanol maser having the smallest average velocity offset of 2.8\,\kms.

\begin{figure}
	\centering
	\begin{tikzpicture}
	
	
	
	\draw \firstellip node [label={[xshift=2.1cm, yshift=-0.9cm]$1720\,\textrm{MHz}$}] {};
	\draw \secondellip node [label={[xshift=2.6cm, yshift=2.1cm]$1667\,\textrm{MHz}$}] {};
	\draw \thirdellip node [label={[xshift=-2.1cm, yshift=-0.9cm]$1612\,\textrm{MHz}$}] {};
	\draw \fourthellip node [label={[xshift=-2.6cm, yshift=2.1cm]$1665\,\textrm{MHz}$}] {};
	\node at (1.5,2.7) {3};
	\node at (-1.5,2.7) {71};
	\node at (0,1.5) {33};
	\node at (0,-0.6) {3};
	\node at (-3,1) {9};
	\node at (3,1) {8};
	\node at (-1.9,1.5) {17};
	\node at (1.9,1.5) {0};
	\node at (-1,0.5) {6};
	\node at (1,0.5) {3};
	\node at (-1.3,-0.8) {0};
	\node at (1.3,-0.8) {11};
	\node at (0,-1.8) {0};
	\node at (0.4,-1.2) {1};
	\node at (-0.4,-1.2) {0};
	
	\end{tikzpicture}
	\caption{A Venn diagram presenting the number of ATLASGAL sources associated with the different transition lines of the hydroxyl maser.} 
	\label{fig:oh_venn_diagram}
\end{figure}
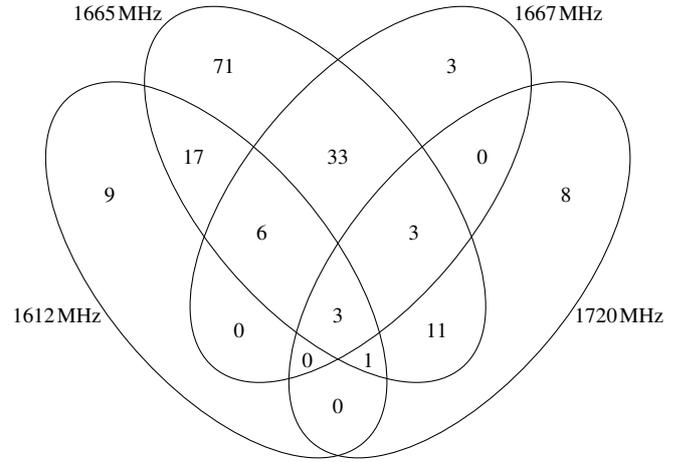

Out of the 807 hydroxyl maser sites identified by the THOR survey, we have matched 294 to 141 individual dense clumps identified by the ATLASGAL survey. This is an association rate of 36\,per\,cent with 2.1 maser sources per clump on average. The majority of clumps contain more than 2 maser sources (75/141), and the remaining 66 sources are only associated with a single maser. Of the 75 clumps with multiple maser associations, there exists 4 clumps with 5 maser sources and also another 4 which contain more than 5 maser sources. As the average number of masers per clumps is 2.1, we find that only 3 of these 8 sources have a significantly\footnote{Significance is based on the 3$\sigma$ value calculated using Poisson statistics.} large number of masers (G030.703-00.067 \& G031.412+00.307 with 9 and G030.823-00.156 with 8; these 3 regions are associated with the W43 star forming complex; \citealt{Motte2002}). For the 495 OH maser sites identified by the SPLASH survey, we have matched 114 of these to 24 individual ATLASGAL sources, giving an average number of maser per clump of 5 and association rate of 23\,per\,cent. Five clumps have a significant number of masers (G338.876-00.084 with 8, G339.986-00.426 with 10, G340.054-00.244 with 12, G340.784-00.097 with 13 and G344.227-00.569 with 8), with the maximum number of masers being 13, associated with G340.784-00.097.

As both surveys cover all four hydroxyl emission lines, we find the 1612, 1665, 1667 and 1720\,MHz transitions to have corresponding association rates with 165 dust clumps of 3, 60, 42, 51\,per\,cent respectively. Figure\,\ref{fig:oh_venn_diagram} presents a Venn diagram of the matches between clumps and the four different hydroxyl transition lines. We find that 145 of the 165 dust continuum sources are associated with a 1665\,MHz maser, with only 26 dense clumps containing a 1720\,MHz maser, showing that the 1665\,MHz are more commonly found towards regions of star formation. The mean association rates for the 1665, 1667 and 1720\,MHz are all roughly 50\,per\,cent, however, the 1612\,MHz has an association rate of 3\,per\,cent. The lower number of associated 1612\,MHz masers is expected due this maser transition being larger associated with the expanding shells of AGB stars. Furthermore, we find the majority of hydroxyl masers that are associated with a clump (96\,per\,cent) to be associated with either a YSO or \hii\ region as identified by the ATLASGAL survey. We only find 8 clumps that are coincident with a hydroxyl maser (1665 or 1667\,MHz) and that are associated with a protostellar object (identified by a weak mid-infrared counterpart in the ATLASGAL survey).

\begin{figure}
	\includegraphics[width=0.49\textwidth]{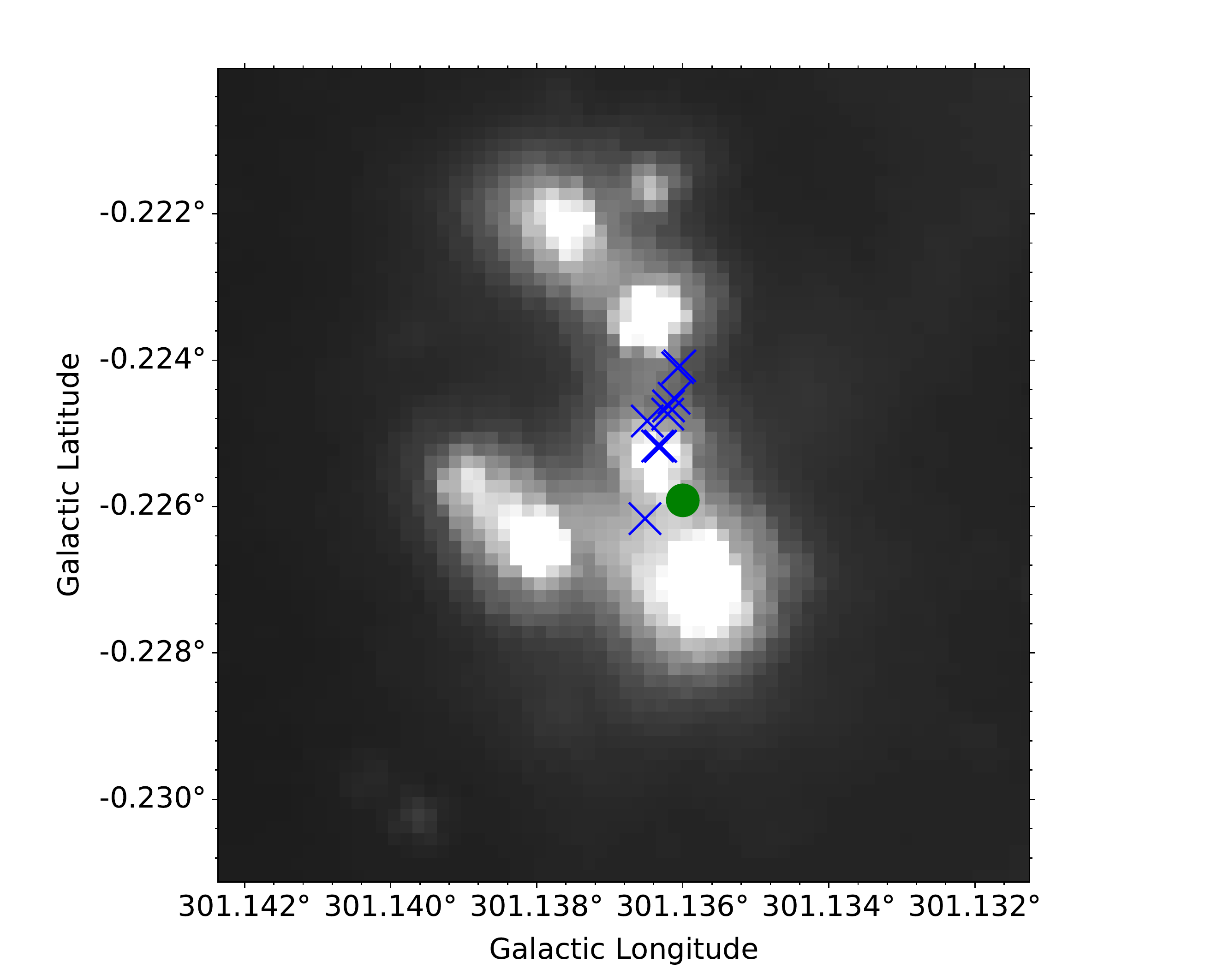}
	\caption{An example GLIMPSE 8\,\micron\ image of the region \mbox{G301.136-00.226}. The positions of water masers associated with this region have been overplotted as blue crosses and the associated 6.7\,GHz methanol masers as green filled circles. It is likely that these masers are associated with a known protostellar outflow in the region (see text for details).}
	\label{fig:example_ir}
\end{figure}

We have matched 345 water maser sites (978 maser spots), presented in \cite{Walsh2014}, to 291 dust continuum sources, an association rate of 55\,per\,cent with a mean maser number of spots per clump of 1.2, which is lower than the overall rate found for the hydroxyl masers. We find that out of the 291 matched dust continuum sources, there is one source with 10 maser spots (G301.136-00.226), a significant number of maser spots, which is also associated with a 6.7\,GHz methanol maser. A 8\micron\ infrared GLIMPSE legacy survey \citep{Churchwell2009} image of this region is presented in Fig.\,\ref{fig:example_ir}.  A linear alignment between the water masers can be seen, likely due to an outflow originating in the region \citep{Henning2000,Guzman2012}. This is consistent with the idea that the morphology of water masers trace outflows (e.g. \citealt{Claussen1996}).

For the 432 emission sites of the 12.2\,GHz, as identified by \cite{Breen2012,Breen2012b,Breen2014a,Breen2016a}, we find that 340 are associated with 330 ATLASGAL sources. The majority of clumps (320; 96\,per\,cent) that are associated with a 12.2\,GHz methanol maser only host a single maser, with the remaining clumps (10; 3\,per\,cent) hosting two masers. No clumps are associated with more than two 12.2\,GHz methanol masers. Matching statistics for the 6.7\,GHz methanol masers are taken unchanged from \cite{Billington2019a}. This study produced a sample of 839 6.7\,GHz methanol masers (identified by the MMB survey) that are associated with dust continuum emission, also with an average of 1 maser site per clump. The percentage of the number of dust clumps associated with each maser species can be found in Table\,\ref{table:maser_matches}.

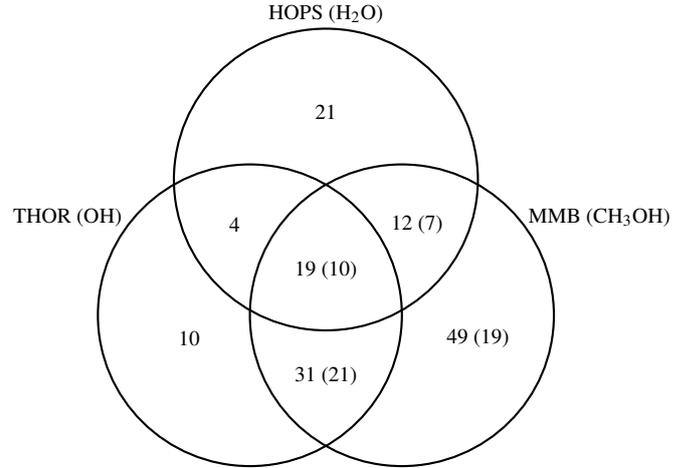
\begin{figure}
	\centering
	\begin{tikzpicture}
	
	\draw[black, thick] (-1,-1) circle (2cm);
	\node at (-3.4,0.3) {THOR (OH)};
	\node at (-1.8,-1.3) {10};
	
	\draw[black, thick] (1,-1) circle (2cm);
	\node at (3.6,0.3) {MMB (CH$_3$OH)};
	\node at (2,-1.3) {49 (19)};
	
	\draw[black, thick] (0,0.8) circle (2cm);
	\node at (0, 3) {HOPS (H$_2$O)};
	\node at (0, 1.7) {21};
	
	\node at (0,-0.4) {19 (10)};
	\node at (-1.2, 0.2) {4};
	\node at (1.2, 0.2) {12 (7)};
	\node at (0, -1.8) {31 (21)};

	\end{tikzpicture}
	\caption{A Venn diagram presenting the number of ATLASGAL sources and the corresponding associated masers between Galactic longitudes, $14.5\degr < \ell < 30\degr$, and Galactic latitudes, $|b| < 0.5\degr$. The values in brackets are the values for the 12.2\,GHz methanol masers, all of which are associated with a 6.7\,GHz maser counterpart.}
	\label{fig:window_venn_diagram}
\end{figure}

\subsection{Common Spatial Area}
\label{sect:window}

Each of the surveys presented in this study have differing Galactic coverages and while this has no impact on the fractional associations between individual masers species and the dust emission, to be able to compare associations rates for different sources, we need to investigate a common region. Fortunately, due to the coverages of the presented surveys, there exists two windows on the Galactic plane where all four maser species were observed in at least one survey. The first window is positioned between Galactic longitudes of $14.5\degr < \ell < 30\degr$ and Galactic latitudes of $|b| < 0.5\degr$, and the second is between Galactic longitudes of $334\degr < \ell < 335\degr$ and Galactic latitudes of $|b| < 0.5\degr$(see Fig.\,\ref{fig:survey_coverages}). Due to the superior sensitivity of the THOR survey, we have chosen to use this first common spatial area to investigate the relationships between the maser species themselves and how they relate to the dust continuum.

Within this window we find that there are 1\,263 ATLASGAL sources, however, only 146 (12\,per\,cent) of these are associated with maser emission. This low association rate may suggest that masers are not always present within clumps or they may only be associated with a particular stage of star formation. The matches between this subsample of clumps and the various maser transitions are shown in Fig.\,\ref{fig:window_venn_diagram} in the form of a Venn diagram. As all of the 12.2\,GHz masers will have a 6.7\,GHz counterpart, due to the nature of the observations, they are included in brackets with the 6.7\,GHz matching values. We find that within this overlapping coverage window there are 64, 56, 57 and 111 clumps associated with hydroxyl (at least one transition), water, 12.2\,GHz methanol and 6.7\,GHz methanol masers respectively.

Out of the 146 maser associated clumps in this window, 47$\pm$4\,per\,cent of them are associated with multiple maser species. We find that 76$\pm$4\,per\,cent of the clumps are associated with a 6.7\,GHz methanol maser, significantly higher than the number of clumps associated with water (38$\pm$4\,per\,cent) and hydroxyl masers (44$\pm$4\,per\,cent). It can also be seen that 12.2\,GHz methanol masers are present in 39$\pm$4\,per\,cent of clumps, showing that only approximately half of 6.7\,GHz maser have a 12.2\,GHz counterpart. This result was found in the 12.2\,GHz maser follow-up catalogues (e.g. \citealt{Breen2012}) but which differs from previous studies of this association rate, e.g. \cite{Blaszkiewicz2004}, who found a rate of 19\,per\,cent. This is likely due to the improved sensitivity of the \cite{Breen2012,Breen2012b,Breen2014a,Breen2016a} observations and the relative dimness of 12.2\,GHz methanol masers compared to the 6.7\,GHz methanol masers. The 6.7\,GHz methanol maser also account for the majority of masers that are found in isolation at 58$\pm$4\,per\,cent, while the water and hydroxyl masers are only found in isolation (no association with other masers in the same clump) 25$\pm$4\,per\,cent and 12$\pm$4\,per\,cent of the time respectively. 

It appears to be quite rare to find only a water maser and hydroxyl maser to be associated with the same source (only 4 out of 146 clumps), whereas the majority of these masers are always found with a methanol maser counterpart. \cite{Beuther2002} used a sample of 29 star forming regions and found that $\sim$62\,per\,cent of 6.7\,GHz methanol masers are associated with water emission and that $\sim$65\,per\,cent of water masers are associated with a 6.7\,GHz methanol maser. Our results differ slightly in that we find only 28$\pm$6\,per\,cent 6.7\,GHz methanol masers are associated with water emission, and 55$\pm$7\,per\,cent of water masers are associated with 6.7\,GHz methanol emission. \cite{Breen2018} presented a comparison between the MMB and HOPS catalogues, and while they also found that 28\,per\,cent of methanol masers have a water counterpart, they only found that 40\,per\,cent of water masers are seen coincident with methanol masers (both 6.7\,GHz and 12.2\,GHz). Although, this disparity is likely due to the matching parameters used in each study and their associated observing sensitivities. Furthermore, \cite{Titmarsh2014} and \cite {Titmarsh2016} conducted studies towards known positions of 324 6.7\,GHz methanol masers and found that $\sim$46\,per\,cent and $\sim$50\,per\,cent of the 6.7\,GHz methanol masers have a water emission counterpart, respectively. The differences in the association rates between the water and methanol masers is likely due to the differing sensitivities for each set of observations and the intrinsic variability of water masers (due to accretion and outflows associated within young stars and protostellar objects). However, the differences could also be due to the smaller sample sizes in previous studies.

\begin{figure*}
	\includegraphics[width=0.49\textwidth]{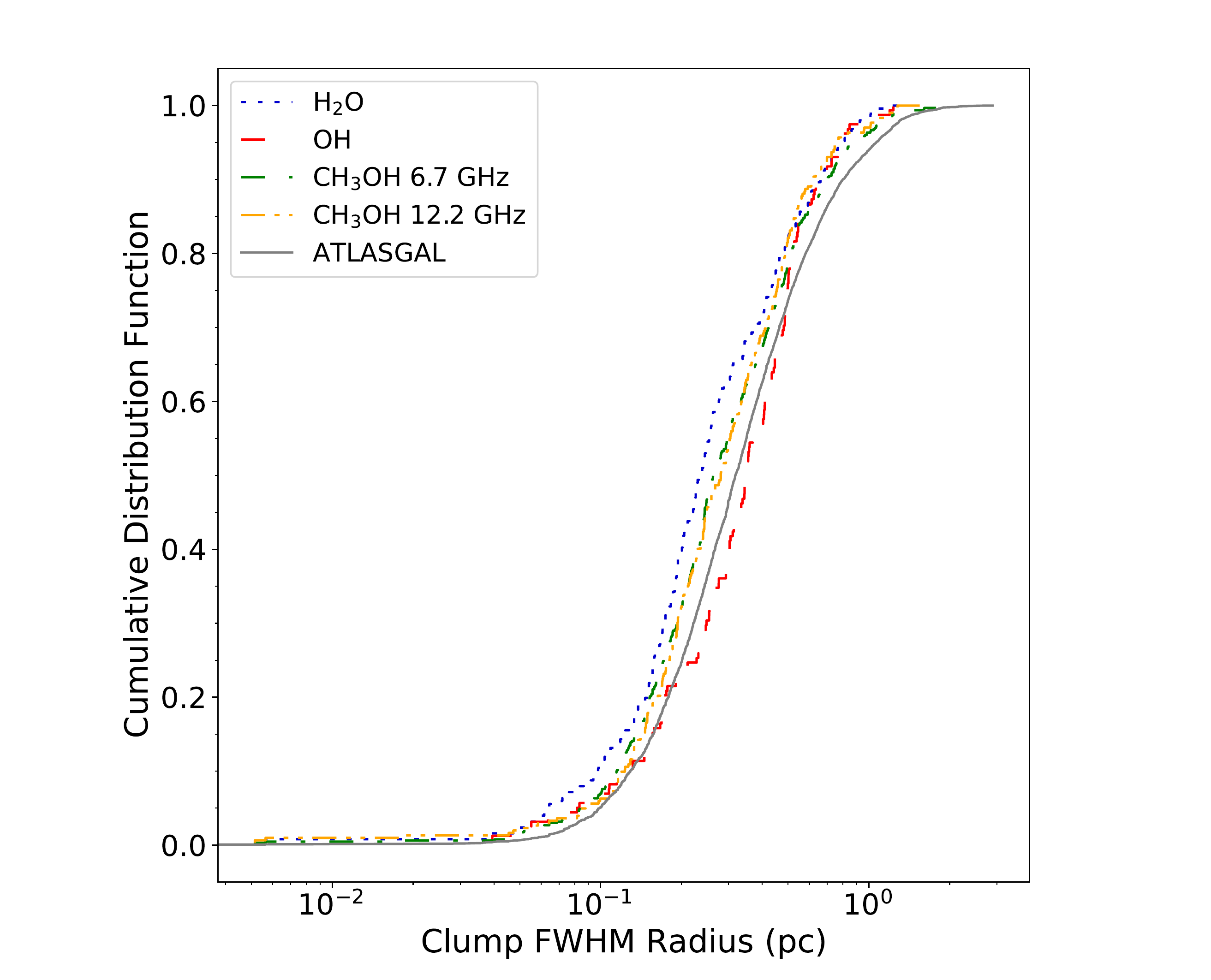} 
	\includegraphics[width=0.49\textwidth]{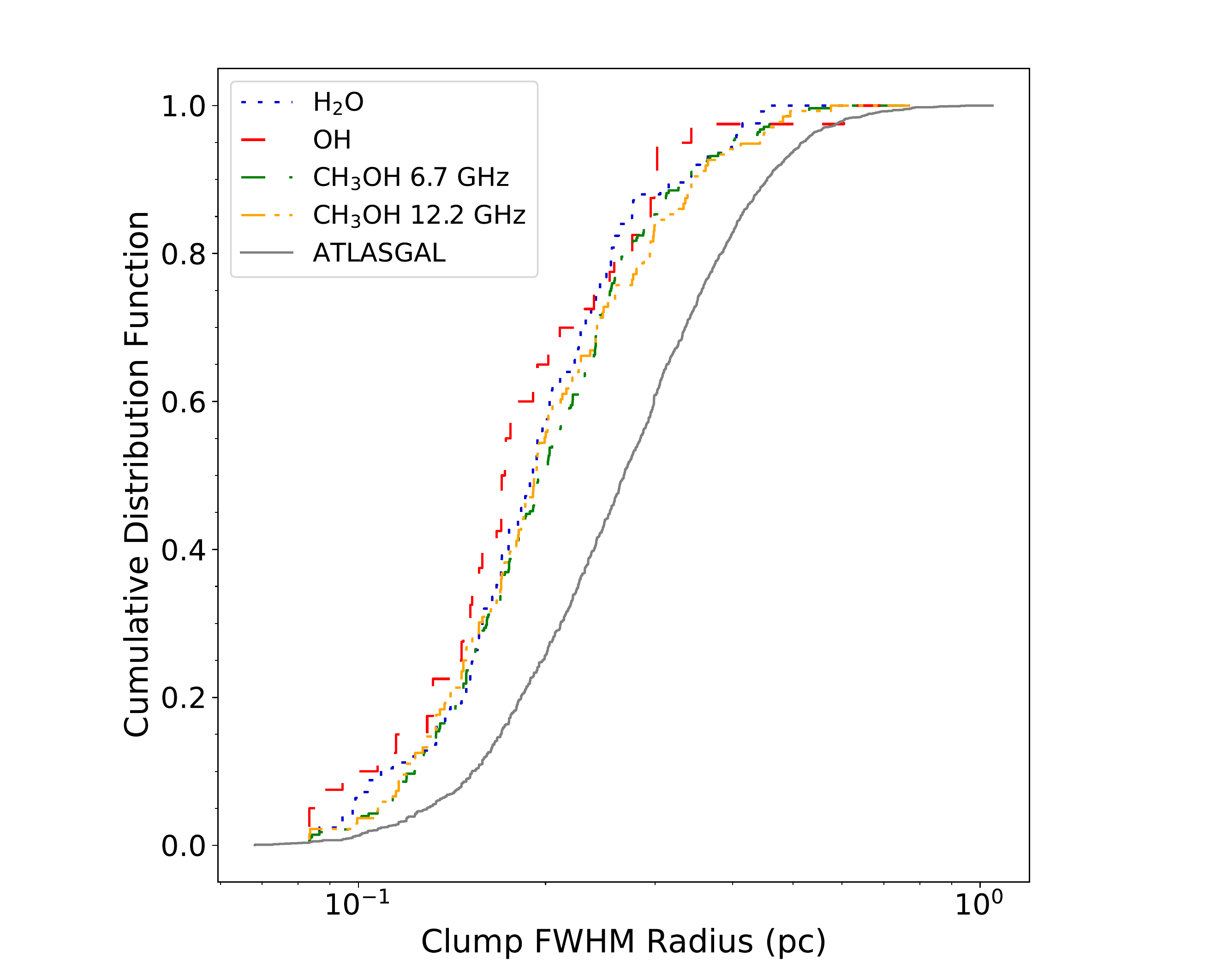} \\
	
	\includegraphics[width=0.49\textwidth]{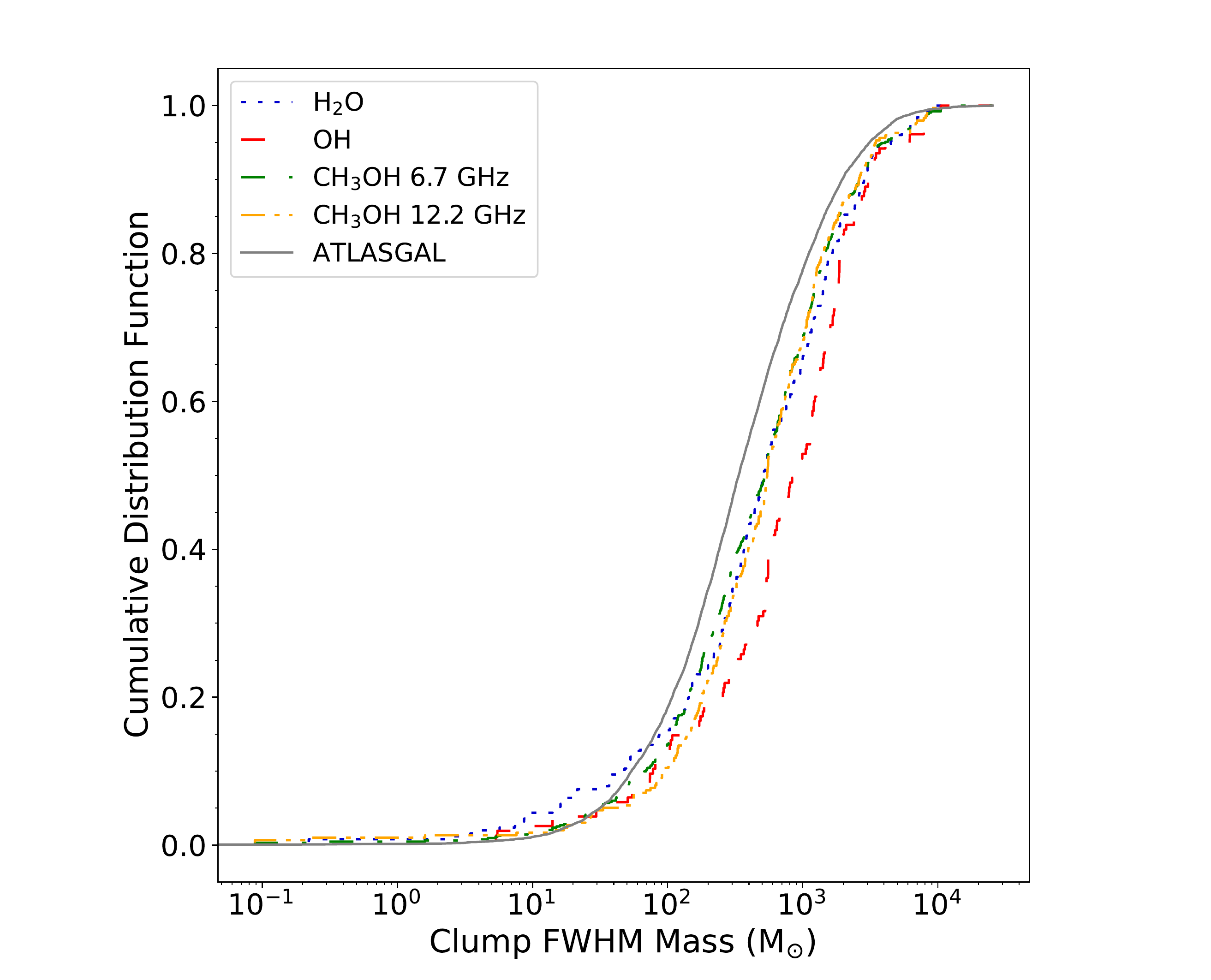} 
	\includegraphics[width=0.49\textwidth]{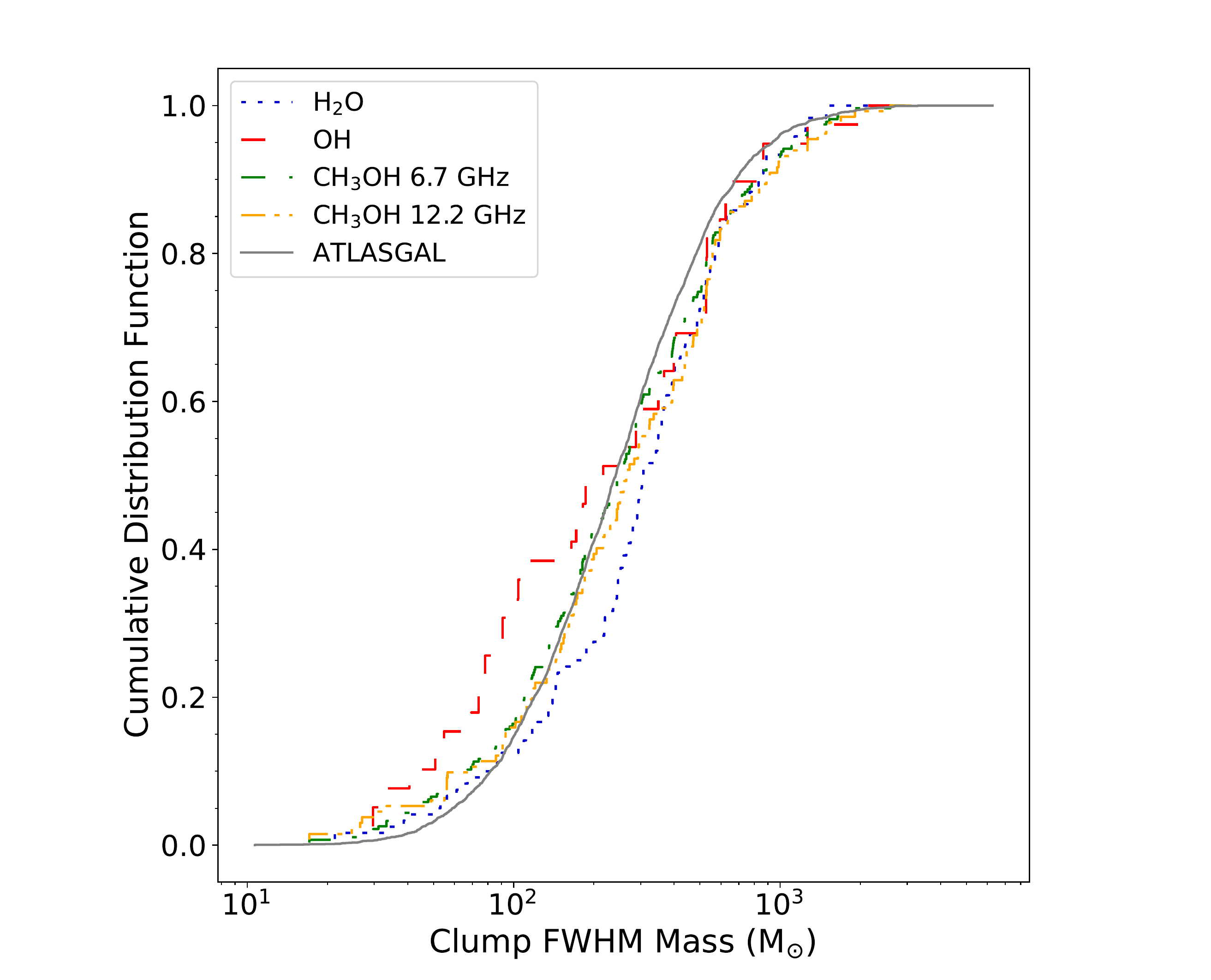} \\
	
	\includegraphics[width=0.49\textwidth]{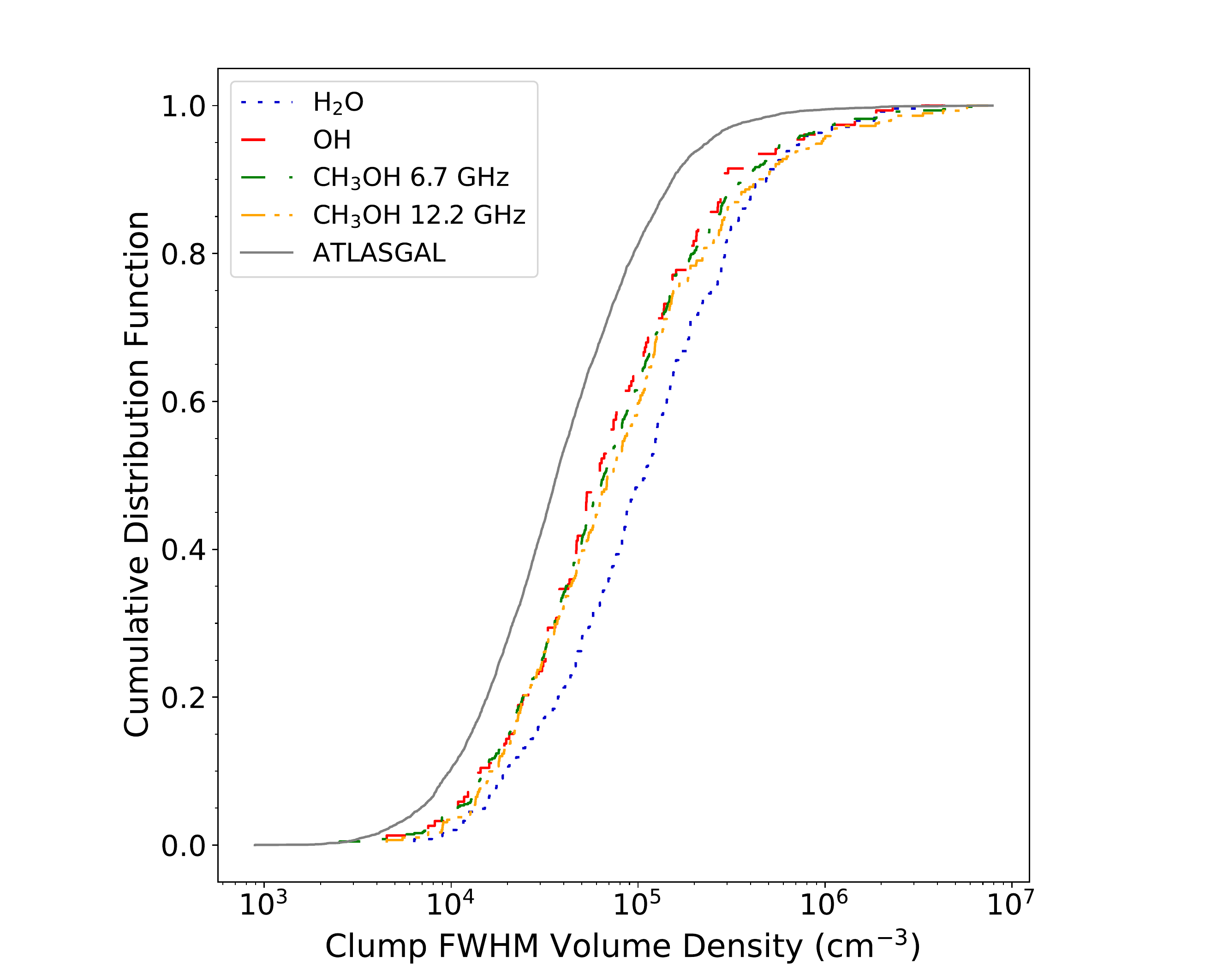}	\includegraphics[width=0.49\textwidth]{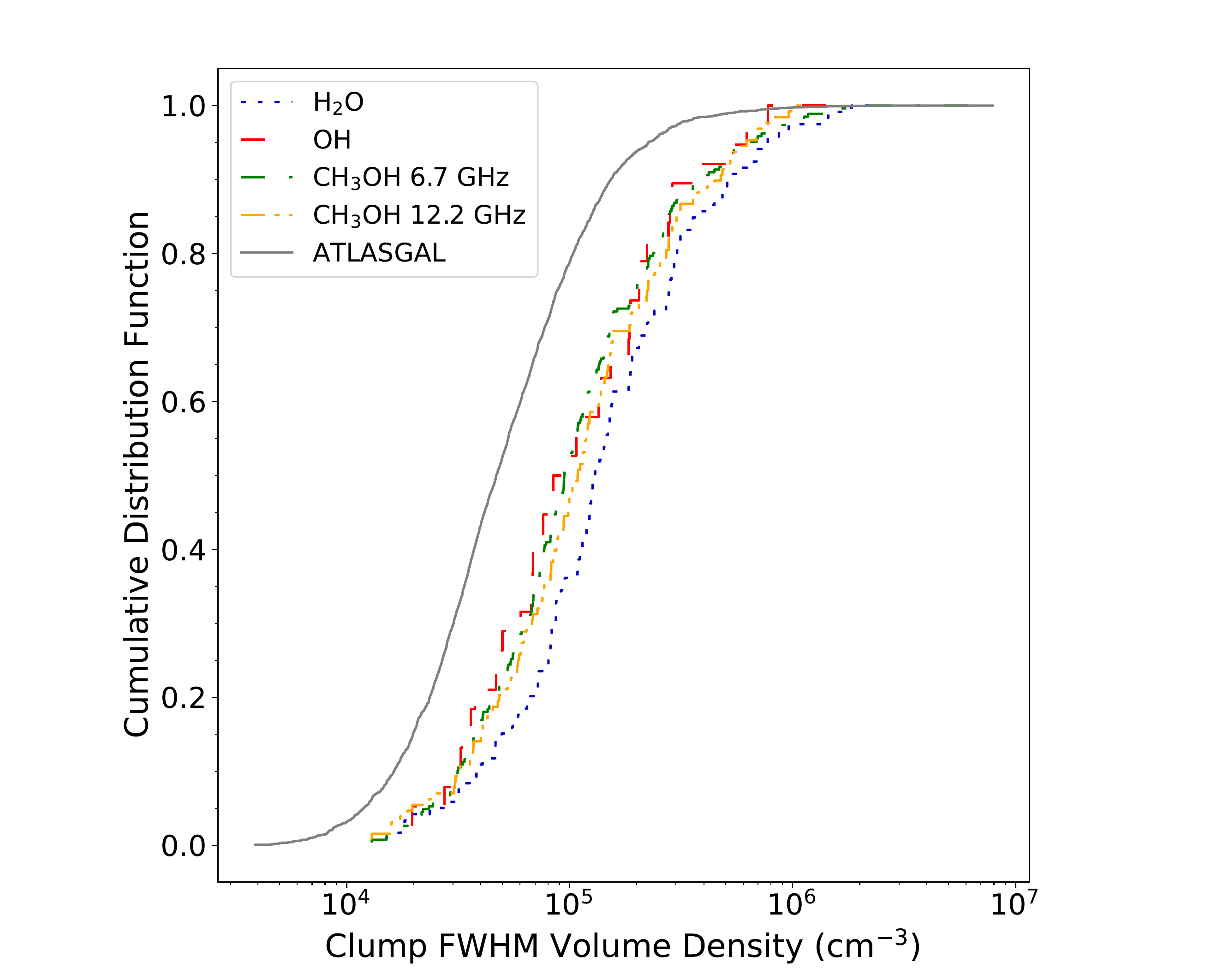} \\
	
	\caption{Clump FWHM Radius, clump FWHM mass and clump FWHM volume density distributions are presented in the upper, middle and lower panels respectively. The cumulative distribution functions in the left panels present the entire distributions of the maser associated clumps and the full ATLASGAL sample, whereas the cumulative distribution functions in the right panels present a distance limited sample (2 to 5\,kpc) of the same distributions. Legends for each sample presented are shown in the top left corner of each panel.}
	\label{fig:clump_radius_mass_density}
\end{figure*}

\begin{figure}
	\includegraphics[width=0.49\textwidth]{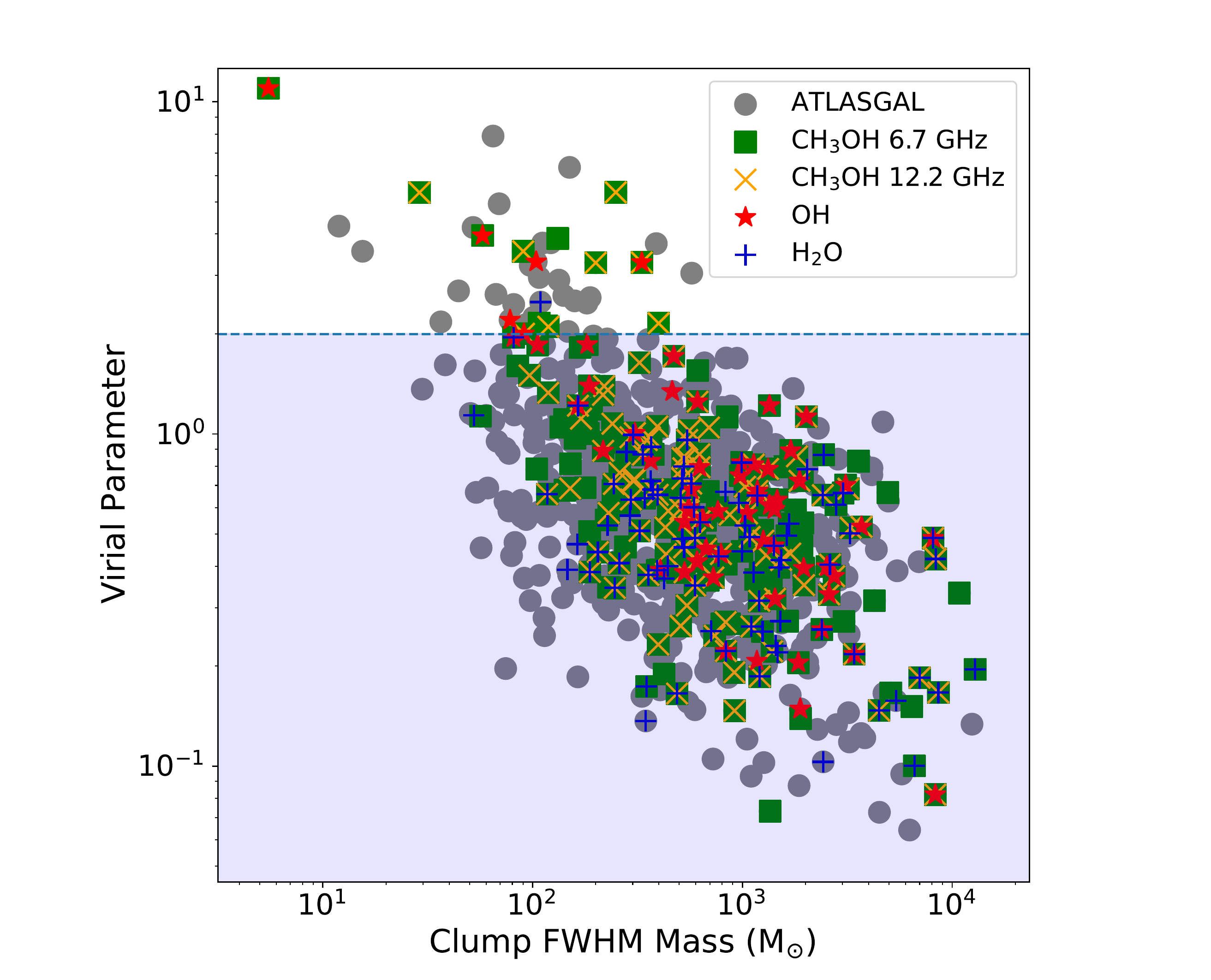}
	\caption{Virial parameters versus clump mass for maser associated clumps. This plot presents a distance limited sample (2 to 5\,kpc). The blue shaded area indicates the region of parameter space where clumps are gravitationally unstable (virial parameter < 2; \citealt{Kauffmann2013a}).}
	\label{fig:virial_parameters}
\end{figure}

\begin{table*}
	\caption{\label{table:complete_stats}Total statistical parameters for all maser associated clumps, determined within our matching criteria, within the Galactic longitude range $5\degr < |\ell| < 60\degr$. The values in brackets show the statistics from a distance limited sample between 2 and 5\,kpc.}
	\begin{tabular}{cccccccc}
		\hline
		\hline
		Parameter & Number & Mean & Standard Deviation & Standard Error & Median & Min & Max \\
		\hline
		
		Distance (kpc) &&&&&&& \\
		\cline{1-1}
		Maser sample & 804 (273) & 5.80 (3.07) & 3.84 (0.48) & 0.14 (0.03) & 4.60 (3.10) & 0.10 (2.10) & 24.20 (3.90) \\                
		ATLASGAL & 7743 (3014) & 4.90 (3.10) & 3.31 (0.46) & 0.04 (0.01) & 3.60 (3.10) & 0.00 (2.10) & 24.20 (3.90) \\
		\hline
		
		Radius (pc) &&&&&&& \\
		\cline{1-1}
		Maser sample & 768 (263) & 0.35 (0.20) & 0.27 (0.08) & 0.01 (0.00) & 0.26 (0.18) & 0.01 (0.08) & 1.85 (0.53) \\                
		ATLASGAL & 5478 (2108) & 0.41 (0.27) & 0.31 (0.11) & 0.00 (0.00) & 0.32 (0.25) & 0.00 (0.07) & 2.89 (0.93) \\
		\hline
		
		Temperature (K) &&&&&&& \\
		\cline{1-1}
		Maser sample & 830 (258) & 23.91 (23.78) & 5.26 (5.66) & 0.18 (0.35) & 23.42 (23.16) & 11.76 (12.13) & 51.23 (51.23) \\                
		ATLASGAL & 9928 (2952) & 19.40 (18.93) & 5.72 (5.89) & 0.06 (0.11) & 18.44 (17.76) & 5.87 (7.89) & 59.96 (51.23) \\
		\hline
		
		log[Luminosity (\lsun)] &&&&&&& \\
		\cline{1-1}
		Maser sample & 780 (258) & 3.99 (3.65) & 1.00 (0.84) & 0.04 (0.05) & 4.00 (3.53) & -0.22 (1.74) & 6.91 (5.83) \\                
		ATLASGAL & 7601 (2952) & 2.95 (2.66) & 1.03 (0.89) & 0.01 (0.02) & 2.89 (2.56) & -0.28 (0.55) & 6.91 (5.93) \\
		\hline
		
		log[Clump Mass (\msun)] &&&&&&& \\
		\cline{1-1}
		Maser sample & 768 (256) & 2.63 (2.30) & 0.68 (0.41) & 0.02 (0.03) & 2.71 (2.32) & -1.05 (1.23) & 4.40 (3.28) \\                
		ATLASGAL & 7460 (2909) & 2.52 (2.32) & 0.61 (0.34) & 0.01 (0.01) & 2.53 (2.33) & -1.05 (1.03) & 4.40 (3.43) \\
		\hline
		
		log[L/M Ratio (\lsun/\msun)] &&&&&&& \\
		\cline{1-1}
		Maser sample & 768 (256) & 1.38 (1.36) & 0.63 (0.65) & 0.02 (0.04) & 1.35 (1.30) & -0.43 (-0.43) & 3.25 (3.25) \\                
		ATLASGAL & 7460 (2909) & 0.44 (0.34) & 0.89 (0.91) & 0.01 (0.02) & 0.42 (0.29) & -2.40 (-2.40) & 3.30 (3.30) \\
		\hline
		
		log[Volume Density (\cmthree)] &&&&&&& \\
		\cline{1-1}
		Maser sample & 744 (248) & 4.86 (5.06) & 0.54 (0.43) & 0.02 (0.03) & 4.83 (5.02) & 3.40 (4.11) & 6.90 (6.90) \\                
		ATLASGAL & 5354 (2052) & 4.59 (4.72) & 0.48 (0.39) & 0.01 (0.01) & 4.56 (4.70) & 2.95 (3.67) & 6.90 (6.90) \\
		\hline
		
		log[Free-Fall Time (yrs)] &&&&&&& \\
		\cline{1-1}
		Maser sample & 744 (248) & 5.27 (5.17) & 0.27 (0.22) & 0.01 (0.01) & 5.28 (5.19) & 4.25 (4.25) & 6.00 (5.64) \\\                
		ATLASGAL & 5354 (2052) & 5.41 (5.34) & 0.24 (0.20) & 0.00 (0.00) & 5.42 (5.35) & 4.25 (4.25) & 6.22 (5.86) \\
		\hline
	\end{tabular}
\end{table*}

\section{Physical parameter calculations}
\label{sect:parameters}

In the following subsections, we describe how the physical parameters of our sample are derived and present the statistics for each in Table\,\ref{table:complete_stats}. The ATLASGAL catalogue provides an effective radius for each clump larger than the beam size of the survey; this radius is the intensity weighted flux density distribution \citep{Contreras2013} scaled by a value of 2.4, which gives a reasonable approximation of the observed clump size \citep{Rosolowsky2010}. However, it was shown by \cite{Urquhart2018} that evolving clumps appeared to be associated with increasing radius, mass and decreasing volume density measurements. They determined that this was due to an observational bias. As Galactic clumps evolve their average temperature increases, which in turn, results in the extended envelope of these regions becoming bright enough to be detected by the ATLASGAL survey. Therefore, the apparent size and mass of sources appeared to increase (and density decreasing proportionally) during evolution. \cite{Billington2019a} eliminated this bias by only considering the flux from within the full-width half-maximum (FWHM) contour of each 870\,\micron\ source. It was shown that this technique effectively removed the observational bias (see Fig.\,8 of \citealt{Billington2019a}). 

Throughout this study we also use a distance limited sample, from 2 to 5\,kpc, to reduce any distance related biases that may exist within the sample.

\subsection{FWHM radius, clump mass and volume density}

The FWHM radius and mass values used in this study are taken directly from \cite{Billington2019a} and are scaled from the values found in \cite{Urquhart2018} by using the ratio between the FWHM flux and the total flux of each clump. 

The FWHM volume densities have also been calculated using the FWHM radii and FWHM masses in the same way as described in \cite{Billington2019a}:

\begin{equation}
n\left({\rm H_2}\right) = \frac{3 }{4 \pi  }\frac{M_{\rm{fwhm}}}{\mu {\rm m_p} R_{\rm{fwhm,pc}}^3},
\end{equation}

\noindent where $n$(H$_2$) is the hydrogen number density of clumps, $\mu$ is the mean molecular weight of the gas (taken here to be 2.8), and m$_\textrm{p}$ is the mean proton mass. $M_{\rm fwhm}$ and $R_{\rm fwhm}$ are the FWHM mass and FWHM radius respectively. Here the clumps are assumed to be spherical and not extended along the line of sight. Figure\,\ref{fig:clump_radius_mass_density} presents the distributions of the clump FWHM radius, FWHM mass and FWHM volume density in the upper, middle and lower panels respectively. It can be seen from the upper panels of this Figure that maser associated clumps are significantly more compact. Although, as shown in the middle panels, the masses of these clumps are similar to the overall distribution of clump masses within the ATLASGAL survey. As the clumps hosting a maser are smaller than the full ATLASGAL clump sample, but have similar masses, the derived volume densities for maser associated clumps are significantly larger.

\subsection{Stability and free-fall times}
\label{sect:freefall_times}

To test the stability of individual clumps we can derive the virial parameter, which is defined as \citep{Bertoldi1992}:

\begin{equation}
\alpha_{\mathrm{vir}}=\frac{5 \sigma_{v}^{2} R_{\mathrm{fwhm,pc}}}{G M_{\mathrm{fwhm}}},
\end{equation}

\noindent where $\sigma_{v}^{2}$ is the velocity dispersion of the \nhthree inversion transition emission observations (e.g. \citealt{Urquhart2011a,Wienen2012,Wienen2018}) and the other parameters are as previously defined. The virial parameter is a measure of the balance between gravitational collapse and internal energy that can counteract this collapse. A value of less than 2 generally indicates that an individual clump is unstable \citep{Kauffmann2013a} and is likely to be undergoing global gravitational collapse in the absence of any supporting magnetic fields. 

The virial parameter has been calculated for 741 ATLASGAL clumps \citep{Urquhart2018}. Figure\,\ref{fig:virial_parameters} presents the distribution of these virial parameters against clump FWHM mass for all clumps. We find that the majority of our clumps have a value lower than 2 and so are likely to be unstable to gravitational collapse, and no differing trends are seen between the maser samples.

Free-fall timescales can be derived for each clump that have a corresponding FWHM volume density measurement:

\begin{equation}
t_{\rm ff} = \sqrt{\frac{3\pi}{32G\rho}}, 
\end{equation}

\noindent where $\rho$ is the mean volume density of a clump and $G$ is the gravitational constant. The free-fall times for sources detected in ATLASGAL range between $\sim$2$\times$10$^4$ and 2$\times$10$^6$\,yrs. These free-fall times will be used to derive the statistical lifetime for each maser species in Sect.\,\ref{sect:stat_lifetimes}. When calculating the free-fall times of clumps, we have assumed that the dominant force is gravity and have not taken into account any support mechanisms that may impede the global collapse of clumps. However, given that $\sim$90\,per\,cent of ATLASGAL clumps are associated with star formation it is safe to assume they are collapsing locally if not globally \citep{Urquhart2018}.

\begin{figure}
	\includegraphics[width=0.49\textwidth]{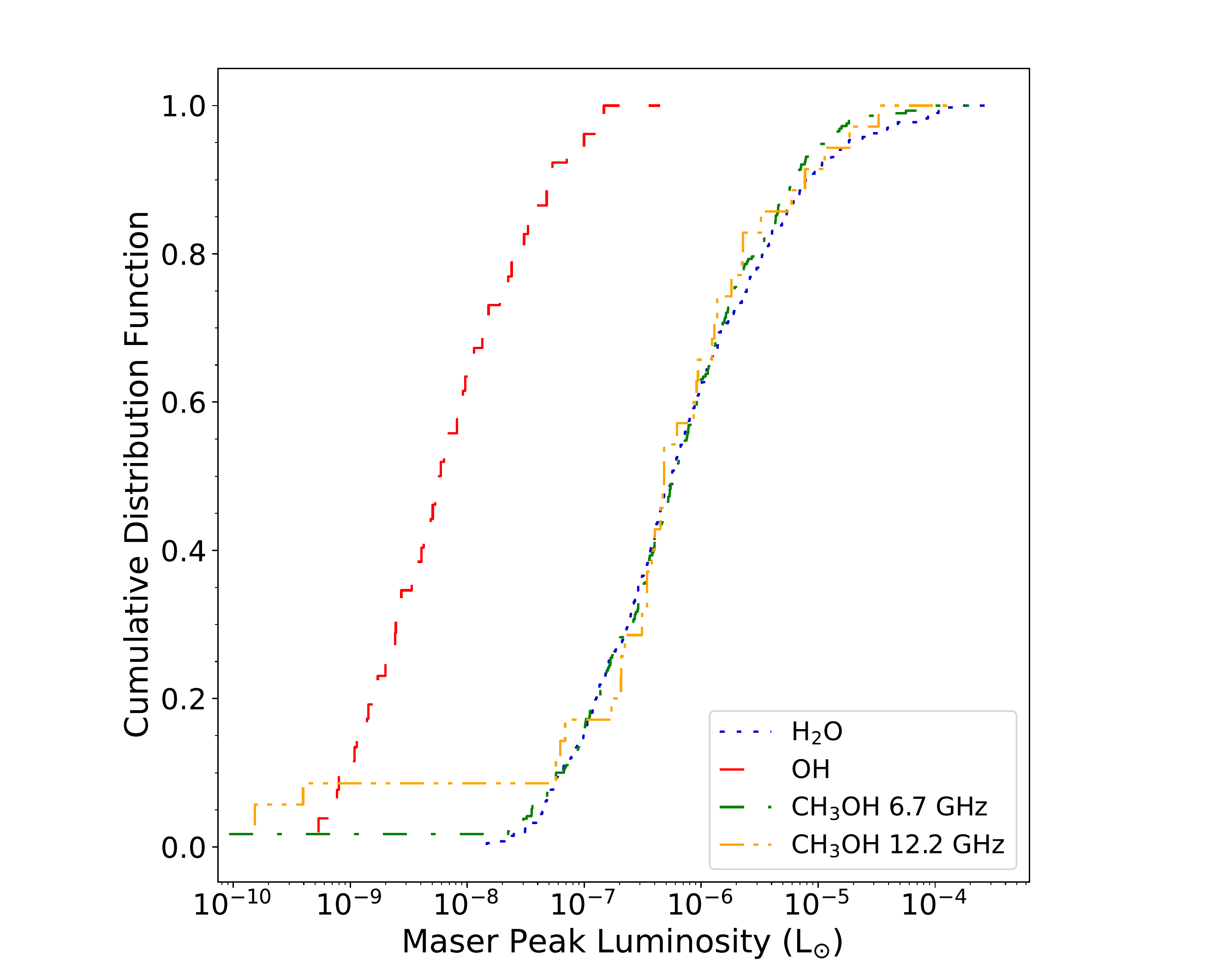}
	\caption{Cumulative distributions function showing a distance limited samples (2 to 5\,kpc) of the maser luminosities for all the species. The legend of each species is given the lower right of the plot.}
	\label{fig:maser_lums}
\end{figure}

\subsection{Maser and Bolometric Luminosities}

\begin{table}
	\caption{\label{table:conversion_factors}Factors for converting the maser luminosities into units of \lsun.}
	\begin{tabular}{cc}
		\hline
		\hline
		Maser Transition Frequency & Conversion Factor (kHz) \\
		\hline
		22.235\,GHz                & 74.1        \\
		1612\,MHz                  & 5.4         \\
		1665\,MHz                  & 5.6         \\
		1667\,MHz                  & 5.6         \\
		1720\,MHz                  & 5.7         \\
		6.7\,GHz                   & 22.3        \\
		12.2\,GHz                  & 40.7        \\
		\hline
	\end{tabular}
\end{table}

Maser luminosities have been calculated using the determined flux density values from the respective maser surveys and distance measurements taken from \cite{Urquhart2018}. For consistency, we have used the peak maser fluxes rather than the integrated fluxes to determine the maser luminosities, as the \cite{Walsh2014} HOPS catalogue does not contain integrated fluxes as they focused on the distribution of individual maser spots. In \cite{Billington2019a}, we show that the differences between the peak and integrated fluxes for maser emission are insignificant above 1\,Jy. The maser luminosities have units of Jy\,kpc$^2$\,\kms\ and are, therefore, somewhat arbitrary. As in \cite{Billington2019a}, we have used conversion factors to convert the maser luminosities into solar units, a list of these factors can be found in Table\,\ref{table:conversion_factors} (see Sect.\,4.5 of \citealt{Billington2019a}). The distribution of maser luminosities can be seen in Fig.\,\ref{fig:maser_lums}. We find that the methanol and water masers have similar luminosities whereas the hydroxyl masers are significantly less luminous. Previous studies have also found similar results (e.g. \citealt{Szymczak2005}). This is expected as it has been theorised that OH masers are associated with the expanding material around \hii\ regions. This material is less dense and so OH maser emission is weaker \citep{Forster1989}.

\begin{figure}
	\includegraphics[width=0.49\textwidth]{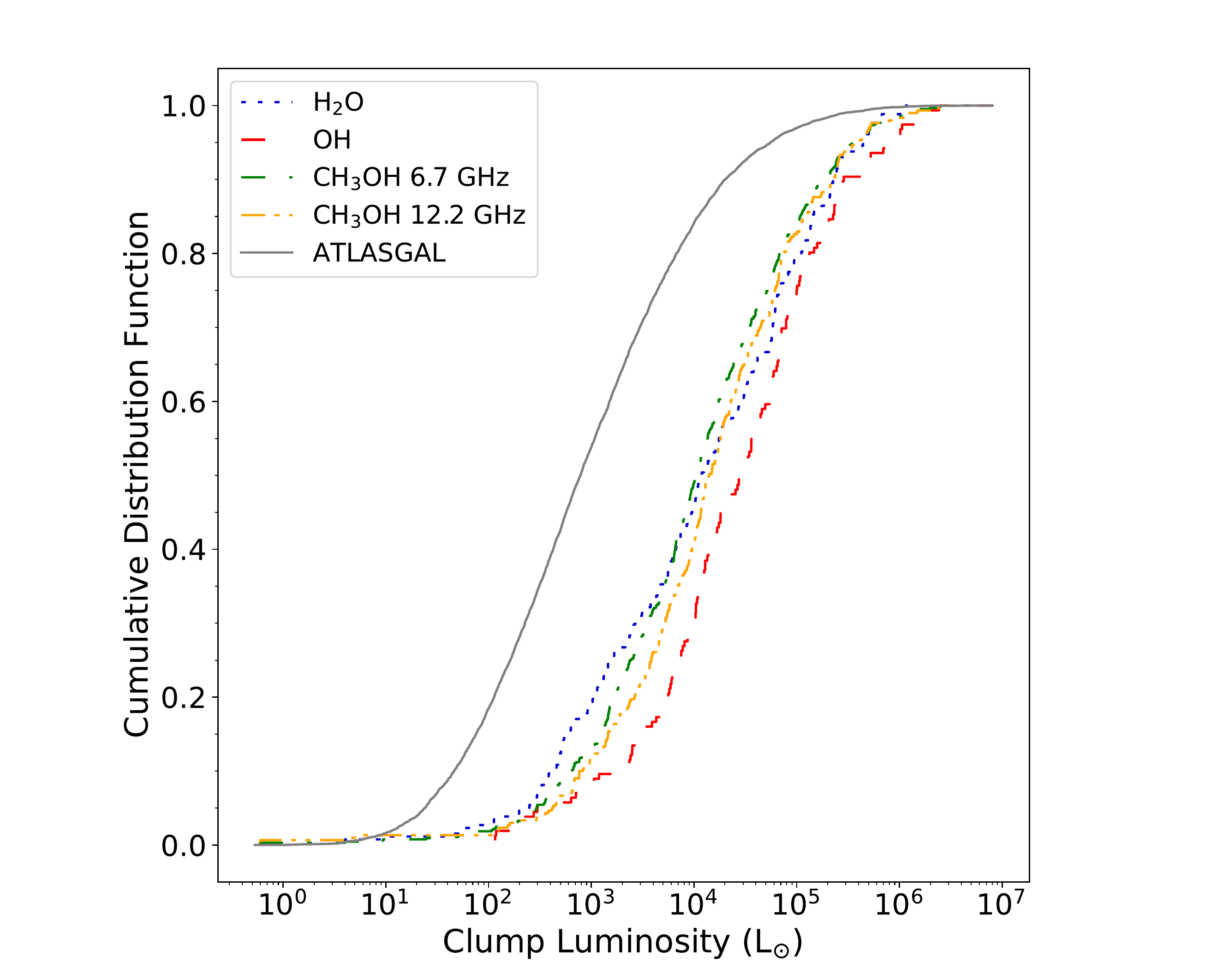} \\
	\includegraphics[width=0.49\textwidth]{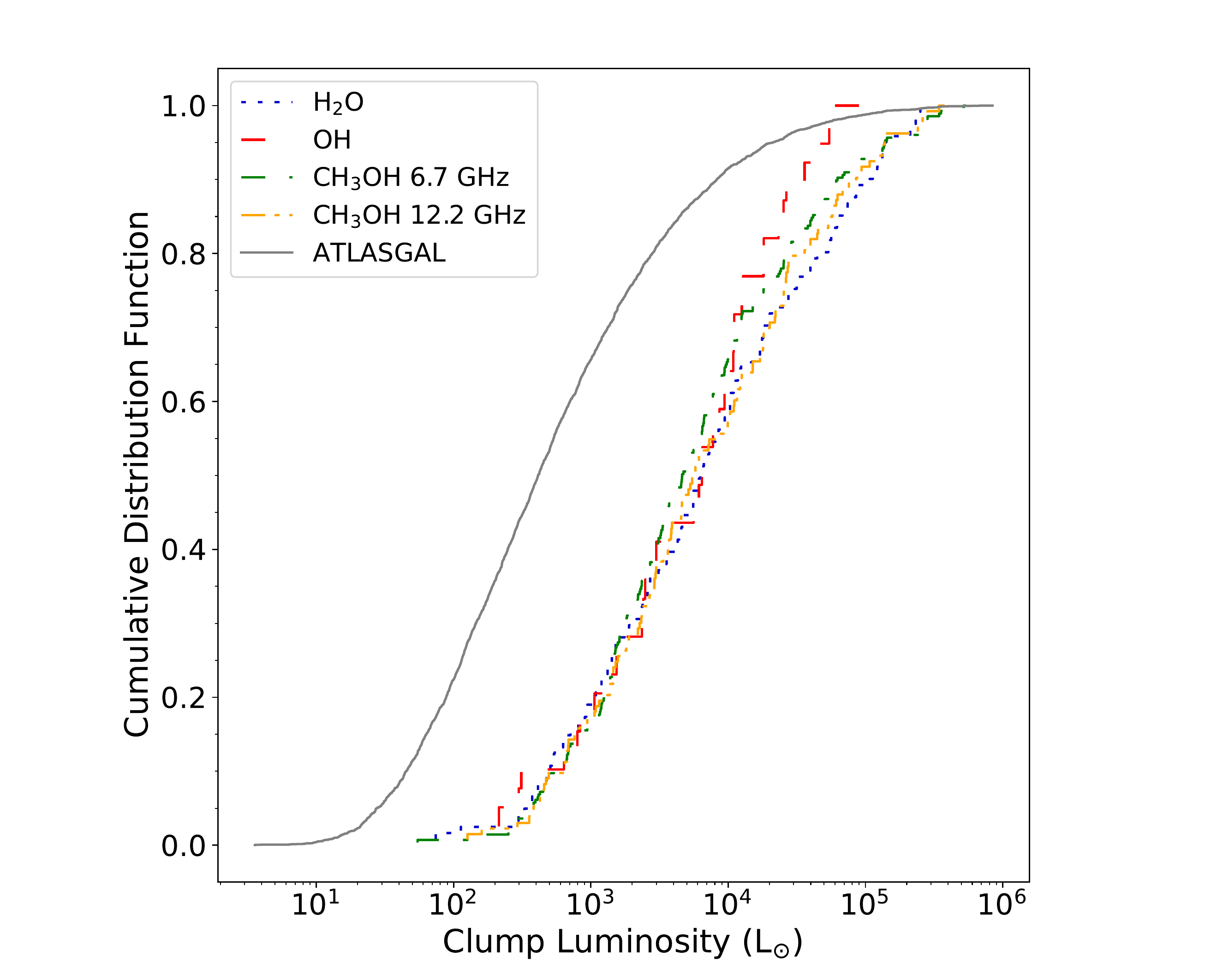} 
	\caption{Clump luminosity parameter distributions. The cumulative distribution function in the upper panel presents the entire distribution of the maser associated clumps and the full ATLASGAL sample. The cumulative distribution function in the lower panel presents a distance limited sample (2 to 5\,kpc). Legends for each maser associated clump are given in the upper top left of both panels.}
	\label{fig:clump_luminosities}
\end{figure}

Bolometric luminosities have been taken unchanged from \cite{Urquhart2018}. These luminosities have been calculated by reconstructing each sources' spectral energy distribution (SED). A more detailed analysis of this method can be found in Sect.\,3 of \cite{Konig2017}. Figure\,\ref{fig:clump_luminosities} presents the distribution of luminosities as cumulative distribution functions (full sample and distance limited sample). There appears to be no significant trends or differences between clumps associated with the different types of masers. Although, they do appear to be significantly more luminous when compared to the full sample of clumps; this will be discussed in detail in Sect.\,\ref{sect:physical_parameters}.

\begin{figure}
	\includegraphics[width=0.49\textwidth]{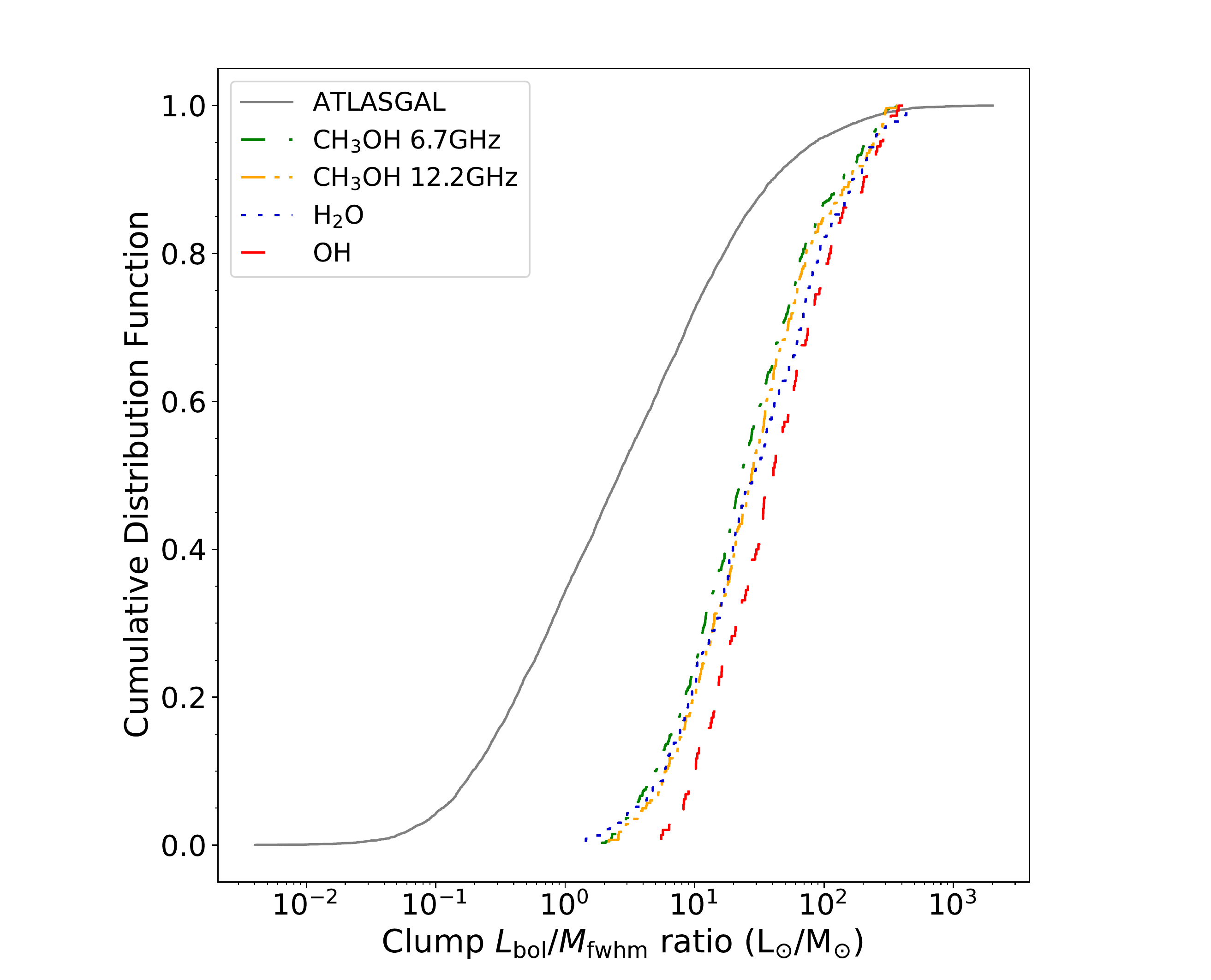}
	\caption{The cumulative distribution function of the \lmratio\ ratios for all maser associated clumps. A legend is shown in the top left for each of the subsamples.}
	\label{fig:lm_ratios_all}
\end{figure}

\subsection{Temperatures and \lmratio\ ratios}

Along with the physical parameters that have been derived in the previous subsections, we also define the \lmratio\ ratio. This is the ratio between the bolometric luminosity of clumps and their corresponding FWHM mass, a distance independent quantity. Previous works have shown that this ratio is a good indicator of evolutionary stage \citep{Molinari2008, Molinari2019}, and so for the analysis within this study we shall use the \lmratio\ ratios of clumps to determine the global evolutionary phase of these star formation regions. The \lmratio\ ratios for the ATLASGAL sample range from 10$^{-2.4}$ to 10$^{3.3}$\,\lsun/\msun\ with a mean of 10$^{0.4}$\,\lsun/\msun. Figure\,\ref{fig:lm_ratios_all} presents the \lmratio\ ratio distributions as a cumulative distribution function, which will be further discussed in Sect.\,\ref{sect:evo_stage}. 

Temperatures for all clumps have also been utilised from the ATLASGAL survey \citep{Urquhart2018}. Along with the \lmratio\ ratios, temperature is a good indicator of evolution as it has been shown that temperature and \lmratio\ ratios of Galactic clumps are tightly correlated (see Figure 22 of \citealt{Urquhart2018}).

\subsection{Uncertainties in Physical Parameters}

The uncertainties in the distance measurements towards clumps are on the order of $\pm$0.5\,kpc and are estimated from the Bayesian distance algorithm presented in \cite{Reid2016}. The uncertainty for radius is linearly correlated with the distance errors and this uncertainty is $\sim$30\,per\,cent at 1\,kpc but only a few percent at distances larger than 10\,kpc. The mean error found when determining the dust temperature values from the spectral energy distributions is $\sim$10\,per\,cent. The fractional uncertainty for the maser luminosities is roughly $\sqrt{2}$ times the fractional uncertainty in the distance measurements. However, in calculating this quantity we are assuming the maser sources are emitting isotropically and so the uncertainty for these measurements are hard to estimate. As for the physical properties of the clumps, the error in the bolometric luminosities is a factor of a few due to the uncertainty on the bolometric flux values (15$-$50\,per\,cent, depending on wavelength of the observations; see \citealt{Konig2017}) and the distance errors. The uncertainty in the mass and volume density are likely dominated by the uncertainty in the value for the dust absorption coefficient (interpolated from \citealt{Schuller2009} and taken to be 1.85cm$^2$\,g$^{-1}$) and the estimation of the dust-to-gas ratio (taken to be 100). Both of these parameters are poorly constrained and lead to uncertainties of a factor of $\sim$2$-$3. The uncertainties on the \lmratio\ ratios are also a factor of a few, following the errors on the luminosity and mass calculations.

While the uncertainties in the physical parameters may be quite large, they affect the entire sample uniformly and will increase the scatter in the distributions. However, this will still allow for statistical trends to be identified and analysed, especially when considering a distance limited sample. The use of a distance limited sample will reduce biases of mass and size due to sensitivity limitations and the large range of distances to the clumps.

\begin{figure}
	\includegraphics[width=0.49\textwidth]{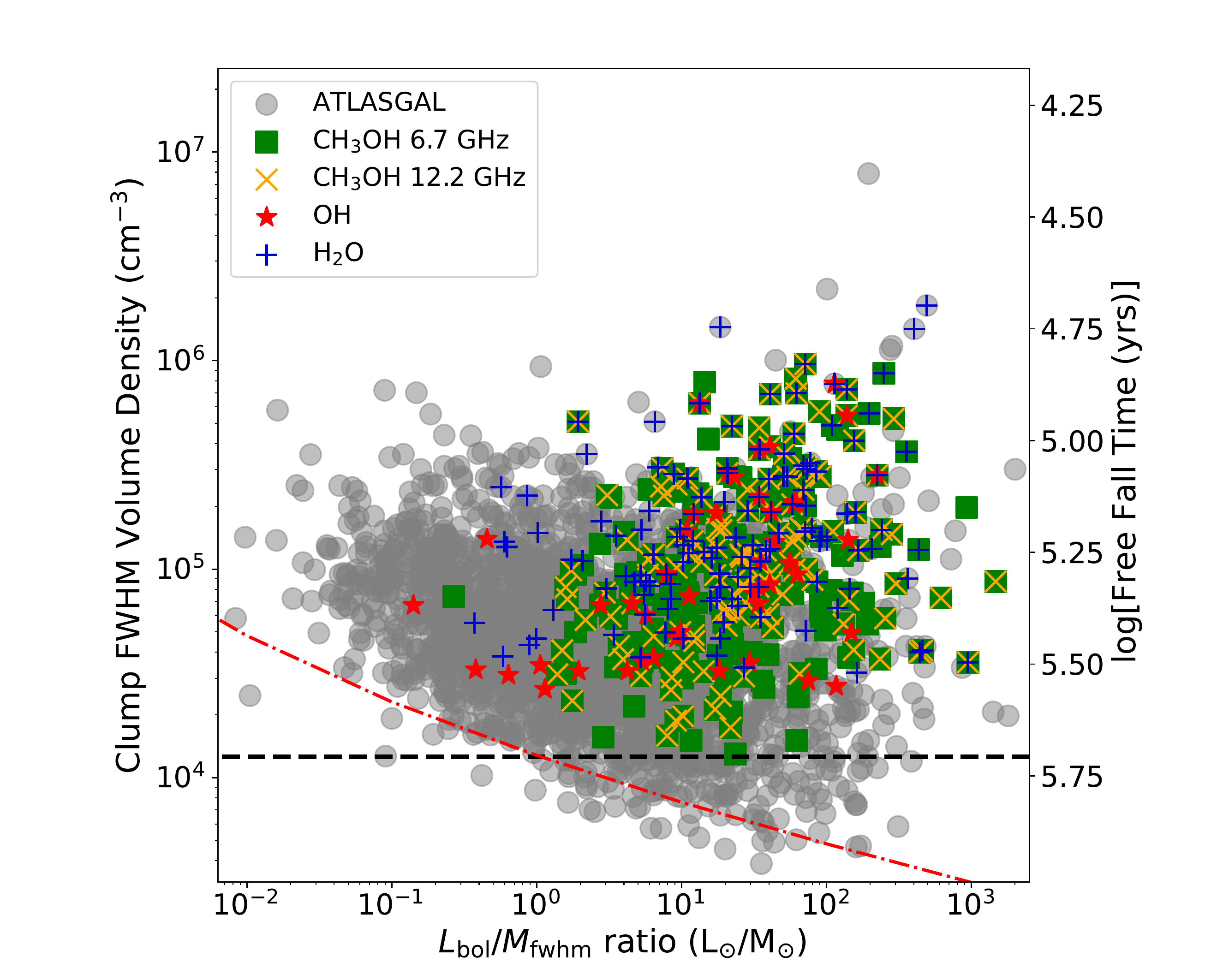}
	\caption{Distance limited sample of volume density versus \lmratio\ ratio for both the maser associated clumps and the full ATLASGAL sample. The entire ATLASGAL sample is shown in grey with the various maser species denoted by the legend in the upper left of the plot. The sensitivity limits of the ATLASGAL survey is shown as the red dash-dotted line, determined for a distance of 2\,kpc. The dotted black line at $y$=10$^{4.1}$\,\cmthree\ is the limit for 6.7\,GHz maser emission to exist, as described in \citealt{Billington2019a}.}
	\label{fig:lm_ratios_vs_vol_den}
\end{figure}

\section{Discussion}
\label{sect:discussion}

As in \cite{Billington2019a} we have opted to only use the central 95\,per\,cent (2$\sigma$) of parameter values within any given distribution. This will remove any potential errors or biases resulting from outliers due to inaccurate distance measurements or extreme sources.

\begin{table*}
	\caption{\label{table:maser_lm_ratios}The central 95\% of the \lmratio\ ratio, volume density (distance limited) and luminosities (distance limited) parameters for each maser species.}
	\begin{tabular}{lccc}
		\hline
		\hline
		Maser Species       & \lmratio\ ratio ranges         & Volume density ranges (\cmthree) & Luminosity ranges (\lsun) \\
		\hline
		Water               & 10$^{0.15}$ $-$ 10$^{2.66}$    & $10^{4.5} - 10^{6.3}$ & $10^{2.6} - 10^{5.4}$  \\
		Hydroxyl            & 10$^{0.74}$ $-$ 10$^{2.59}$    & $10^{4.4} - 10^{5.9}$ & $10^{2.5} - 10^{5.1}$  \\
		Methanol 12.2\,GHz  & 10$^{0.32}$ $-$ 10$^{2.60}$    & $10^{4.2} - 10^{6.0}$ & $10^{2.6} - 10^{5.4}$  \\
		Methanol 6.7\,GHz   & 10$^{0.27}$ $-$ 10$^{2.57}$    & $10^{4.1} - 10^{6.0}$ & $10^{2.5} - 10^{5.4}$  \\
		\hline
	\end{tabular}
\end{table*}

\subsection{Analysis Tool: Kolmogorov-Smirnov Test}

Throughout this work, we employ the use of a two-sample Kolmogorov-Smirnov (KS) test. A two-sample KS test is a non-parametric test that compares the empirical cumulative distribution functions for two samples. The test measures the largest difference between two distributions and the associated confidence value, referred to as a $p$-value. The null hypothesis is that two samples are drawn from the same parent population. However, the null hypothesis can be rejected if the $p$-value is smaller than a selected confidence threshold (taken here as 3$\sigma$; i.e. $p$ < 0.0013). This allows us to conclude that there is sufficient evidence to consider the samples to be drawn from different populations and that two samples are significantly different from each other.

\subsection{Physical Parameters}
\label{sect:physical_parameters}

Figure\,\ref{fig:clump_radius_mass_density} presents the full sample and distance limited sample cumulative distribution functions of clump FWHM radius, FWHM mass and FWHM volume density. It can be seen that the clumps associated with a maser are more compact than the full ATLASGAL sample, as presented in the upper right panel of Fig.\,\ref{fig:clump_radius_mass_density}. The ratio between the mean clump radii for maser clumps versus non-maser clumps is 0.85. A two-sample Kolmogorov-Smirnov (KS) test has also been applied to the radii of clumps associated with a maser source and the ATLASGAL sample to identify whether the difference in radii between the two samples is statistically significant. This has revealed that all of the maser associated clumps are significantly more compact than the rest of the dust clump sample ($p \ll 0.0013$). 

We have also assessed the differences in temperature between clumps hosting a maser and those that do not. As can be seen in Table\,\ref{table:complete_stats}, clumps with associated maser emission are warmer than their non-maser counterparts. The mean temperature of a maser clump is 23.9\,K whereas the mean temperature of an ATLASGAL clump is 19.4\,K. We have also tested the samples using a KS test and find that the differences in temperature between the two samples is statistically significant ($p$ $\ll$ 0.0013). This difference has also been noted in another previous study by \cite{Jones2020}, who examined a sample of 731 MMB sources with compact emission at four Hi-GAL wavelengths, and investigated the association between masers and Hi-GAL sources.

The masses of maser associated clumps are similar to the average mass of dense clumps and we do not find any significant differences, as confirmed by a KS test. As the clumps containing a maser are significantly more compact than the ATLASGAL sample, while having similar masses, the calculated volume densities are naturally increased. This shows that all maser species of interest are only associated with those clumps above a certain density threshold ($n$(H$_2$) > 10$^{4.1}$\,\cmthree); this density limit is consistent with the result found in \cite{Billington2019a} for the 6.7\,GHz methanol masers. 

Figure\,\ref{fig:clump_luminosities} presents the distribution of clump luminosity for our various samples and we find that all of the maser clumps are more luminous by a factor of approximately 10 when compared to the average clump luminosity. The mean and median luminosities for maser associated clumps are both $\sim$10$^4$\,\lsun. These values are $\sim$\,2 times higher than those found by \cite{Jones2020}, likely due to the differences in the way that the luminosity measurements were conducted. This implies that a certain protostellar luminosity is required, and therefore, protostellar mass in order to drive sufficient radiative and mechanical energy into the circumstellar environment to effectively pump the various maser species.

The physical parameters discussed above are all similar for each of the clumps associated with a different maser species, and all appear to be significantly larger when compared to the full ATLASGAL sample, except clump radius. These results do not completely conform with previous studies. For example, \cite{Breen2011} found that water masers are typically found towards clumps with larger radii, increased mass and increased luminosity. While we do find that masers are associated with clumps of increased luminosity, these clumps are found to be more compact than the larger ATLASGAL sample, and also have similar masses. This conflict is likely to be due to the source sample used in \cite{Breen2010}, which originates from \cite{Hill2005}. This millimetre study presented observations of 131 star-forming complexes suspected of harbouring massive star formation. The statistically significant differences between the samples presented in this study and \cite{Hill2005} can be attributed to the method of calculating the radii and masses of individual sources. The radii of millimetre sources in \cite{Hill2005} is highly dependent on the temperature of each region, an observational bias that is described in \cite{Billington2019a}, and instead we chose to calculated the sizes of sources based on the FWHM flux distribution. Moreover, \cite{Hill2005} used a constant temperature of 20\,K to determine the mass of each clump, whereas the ATLASGAL temperatures are based on the results of spectral energy distributions, we therefore consider our results to be more reliable. 

\begin{figure}
	\includegraphics[width=0.47\textwidth]{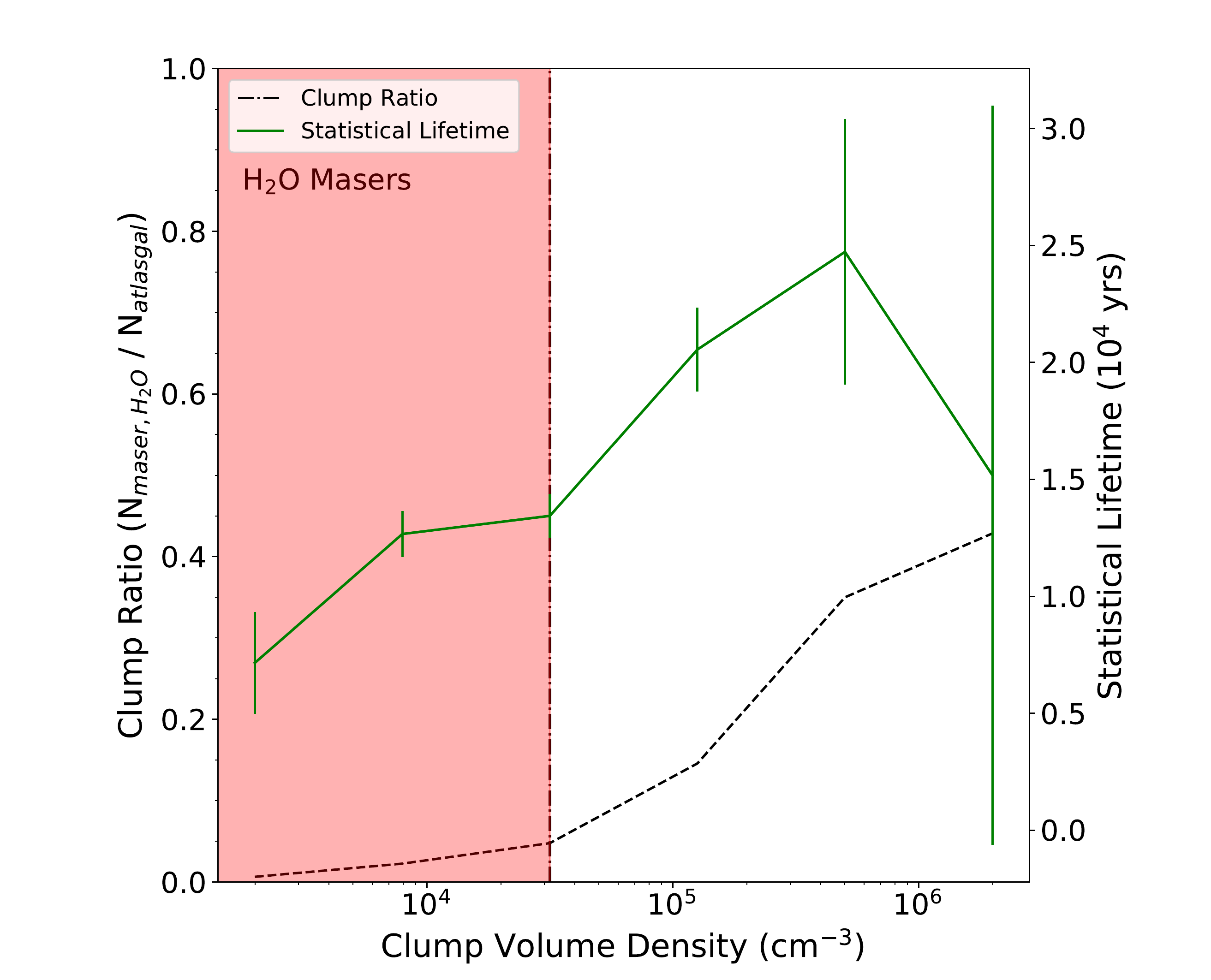} \\
	\includegraphics[width=0.47\textwidth]{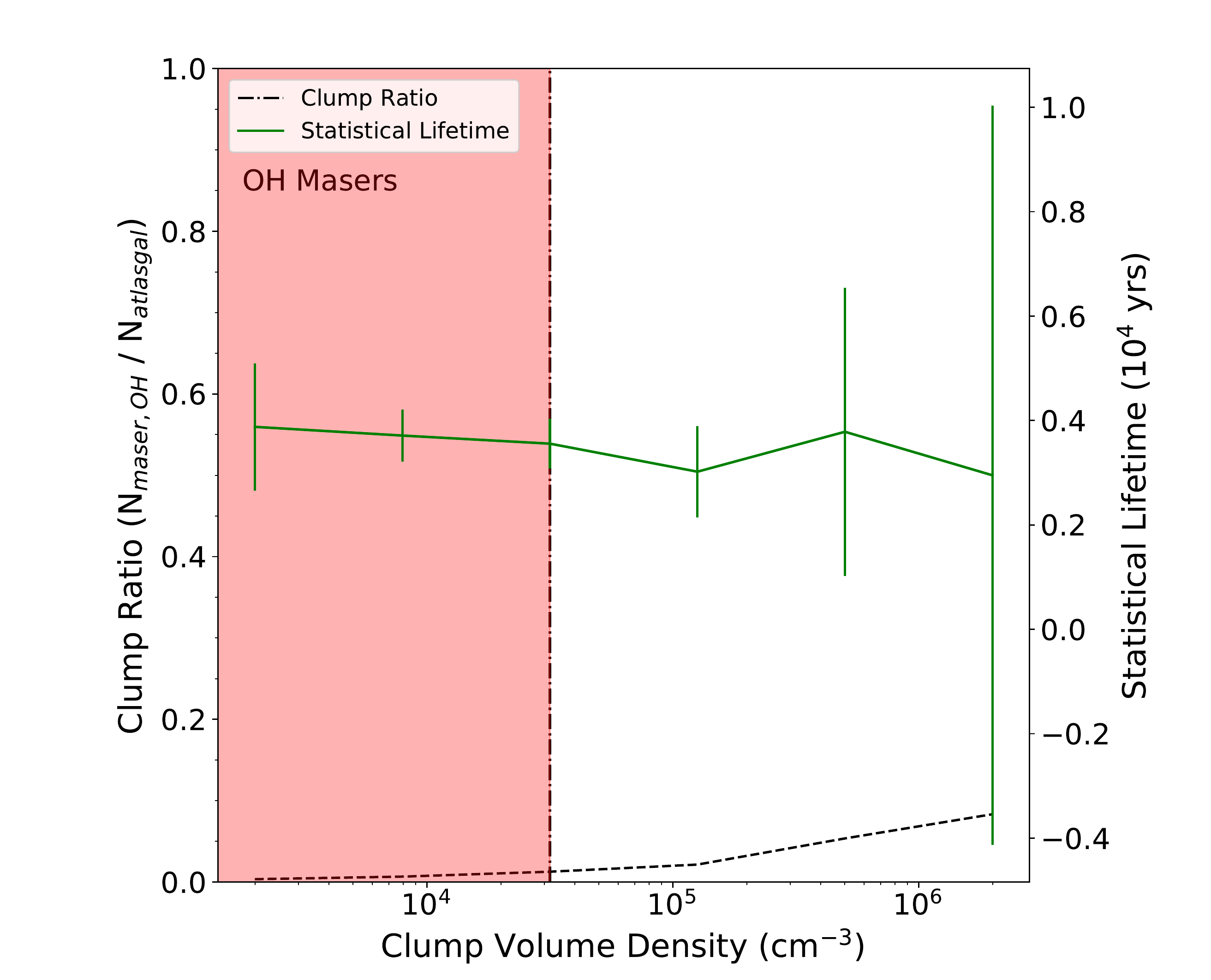} \\
	\includegraphics[width=0.47\textwidth]{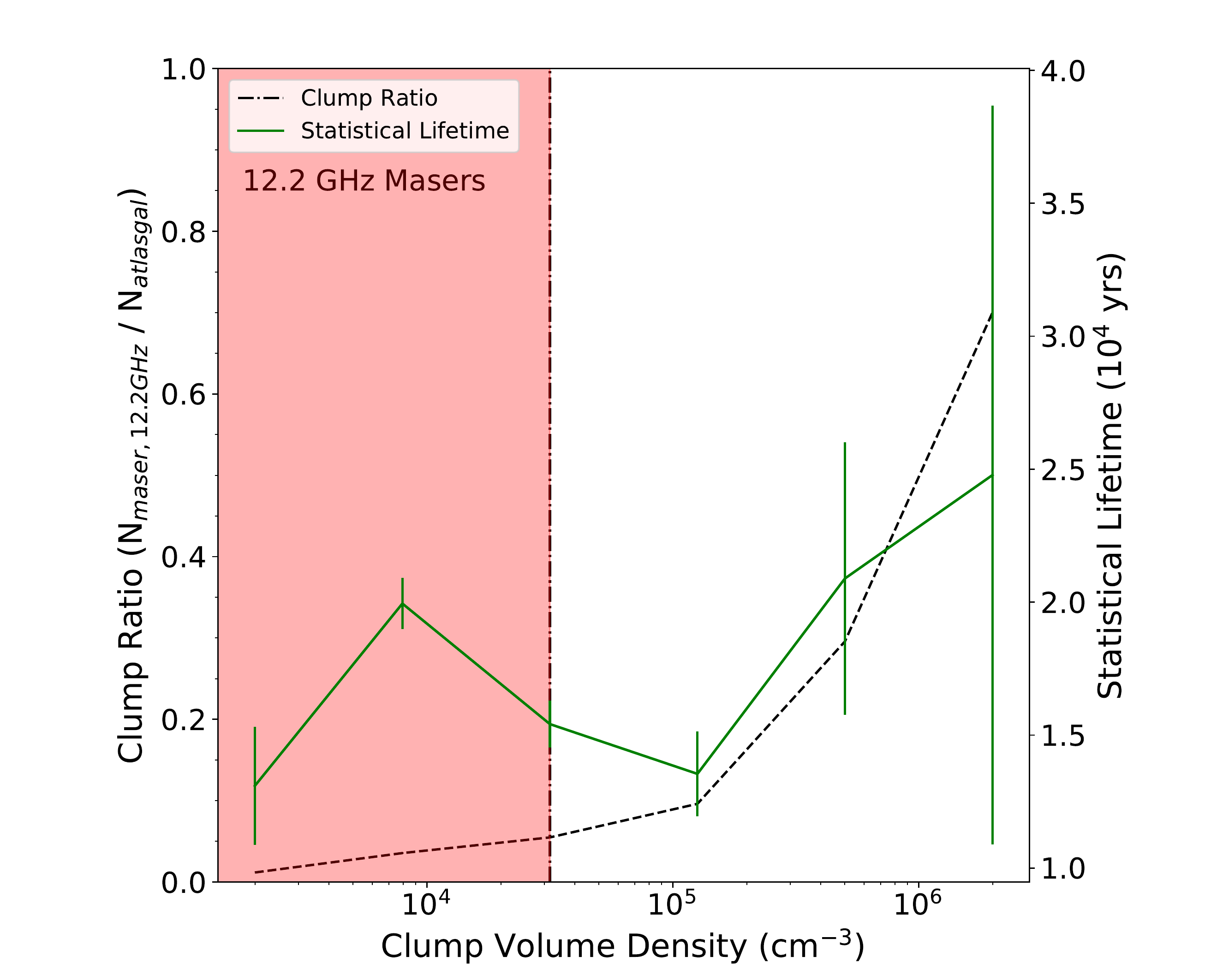}
	\caption{Plots presenting the clump ratios for each maser species and the statistical lifetimes as a function of clump FWHM volume density. The upper, middle and lower panels present the lifetimes for the water, hydroxyl and 12.2\,GHz methanol masers respectively. The black dotted lines shows the increasing number of maser associated clumps with respect to the volume density, whereas the solid green line presents the change in statistical lifetime for each volume density range. The shaded regions represents the parameter space where our sample is incomplete. Errors shown are derived from Poisson statistics.}
	\label{fig:stat_lifetimes}
\end{figure}

\begin{figure*}
	\centering
	\includegraphics[width=\textwidth]{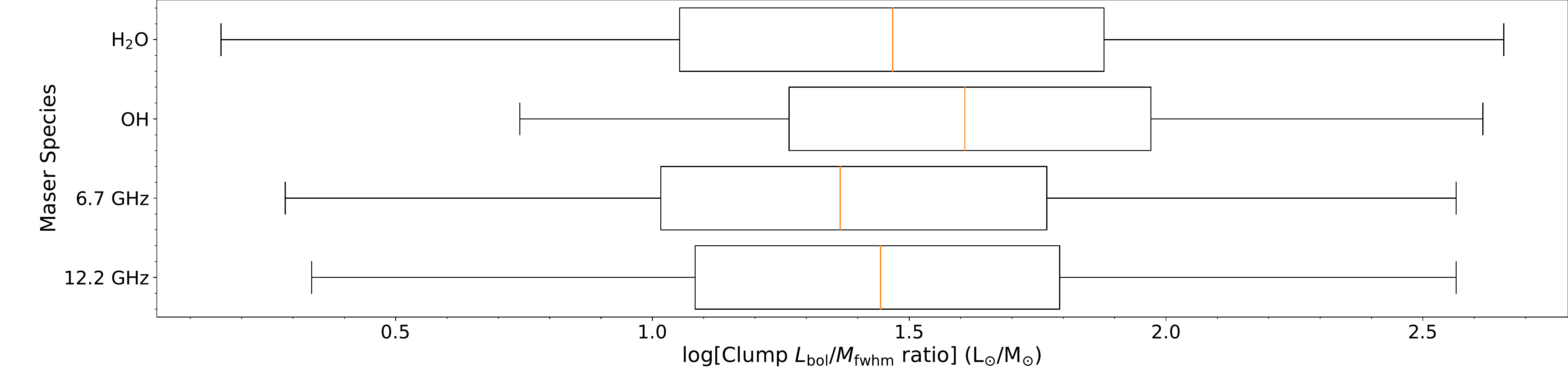}
	\caption{Box plot presenting the central 95\,per\,cent distributions of \lmratio\ ratios for clumps associated with the various maser species. Each box extends from the lower to upper quartile with an orange line denoting the median value, the whiskers shows the full range of the data.}
	\label{fig:evo_stage_boxplot}
\end{figure*}

\subsubsection{Evolutionary Stage - L$_{\rm{bol}}$/M$_{\rm{fwhm}}$ ratios}
\label{sect:evo_stage}

Figure\,\ref{fig:lm_ratios_all} presents the cumulative distribution function for the \lmratio\ ratios of all the maser species and the full ATLASGAL sample. It can be seen from this Figure, that all the maser species occupy approximately the same distinct part of the parameter space, between $\sim$10$^{0.2}$ and 10$^{2.7}$\,\lsun/\msun. The exact ranges of the \lmratio\ ratios for each maser species can be found in Table\,\ref{table:maser_lm_ratios}. 

We have tested the \lmratio\ ratios using clumps that have masses of $> 500$\,\msun\ and by only using the clumps with the smallest offsets to their associated maser sources (spatial offset $<$ 10\,arcsec, velocity offset $<$ 5\,\kms). We find that these restrictions have no effect on the overall distributions of the \lmratio\ ratios for each maser sample, and the corresponding \lmratio\ ratio ranges. Since the \lmratio\ ratios of all of the maser associated clumps are similar (as confirmed by a KS test), this suggests that the mechanisms required for the production of any maser emission only occurs at a set stage in the evolution of star formation and then only for protostellar sources above a certain luminosity ($\sim$500\,\lsun).

In Fig.\,\ref{fig:lm_ratios_vs_vol_den} we show clump FWHM volume density versus clump \lmratio\ ratio. It can be seen that there is a slight negative correlation between these two parameters. Although, this is likely to be due to the sensitivity limit of the ATLASGAL survey, which is also shown in the Figure. In \cite{Billington2019a}, a lower limit of volume density for 6.7\,GHz methanol maser associated clumps was found to be 10$^{4.1}$\,\cmthree. This limit seems to hold for all of the maser species presented in this study, as shown in Fig.\,\ref{fig:lm_ratios_vs_vol_den} and Table\,\ref{table:maser_lm_ratios}. Overall, our results show that certain physical conditions are necessary for the presence of any type of maser. These conditions are as follows: \lmratio\ of between 10$^{0.2}$ and 10$^{2.7}$, volume densities of above 10$^{4.1}$\,\cmthree, and luminosities of above $\sim$500\,\lsun. This luminosity value corresponds to a protostellar mass of $\sim$6\,\msun\ assuming that $L \sim M^{3.5}$ \citep{Kuiper1938} and that the majority of a clump's luminosity arises from a single object. It also must be assumed that for this minimum mass to be accurate, any maser emission emanating from a clump must be associated with the most massive star. This scenario may not always be true, as it has been observed that water masers can arise from low-mass protostars. Naturally, these conditions are only required for masers present in star formation regions and may not hold true for the same maser species associated with other types of celestial objects.

\subsubsection{Statistical lifetimes}
\label{sect:stat_lifetimes}

\cite{Billington2019a} presented the calculation of the statistical lifetime of the 6.7\,GHz methanol maser by finding the ratio of maser associated clumps at specific volume density intervals and multiplying these by the free fall times in these intervals. It was found that the mean statistical lifetime for this maser transition is $\sim$3.3$\times$10$^4$\,yrs, which is in very good agreement with theoretical predictions (\citealt{VanDerWalt2005}; 2.5$-$4.5$\times$10$^4$\,yrs). 

We have repeated this derivation for the maser species presented in this study. Free fall times have been calculated for each clump within our sample that has a corresponding volume density measurement (as discussed in Sect.\,\ref{sect:freefall_times}). These free fall times range from $\sim$20\,000\,yrs to $\sim$750\,000\,yrs. It is likely that any maser emission will only be present for a fraction of these timescales and so by multiplying these times by the fraction of clumps associated with a particular kind of maser emission at specific volume density intervals, the statistical lifetime for maser emission can be found. Figure\,\ref{fig:stat_lifetimes} presents the clump ratios and statistical lifetimes for each maser species as a function of volume density.  It can be seen from this Figure that the number of maser associated clumps increases with increased clump volume density.  For the calculation of the statistical lifetimes, we only include clumps that have volume densities above our completeness limited for this parameter, $n>10^{4.5}$\,\cmthree\ \citep{Billington2019a}. We find that the mean lifetimes of maser emission to be 1.6, 0.4 and 2.0$\times$10$^4$\,yrs for the water, hydroxyl and 12.2\,GHz methanol masers respectively. The statistical lifetime for the 6.7\,GHz maser is taken from \cite{Billington2019a} as 3.3$\times$10$^{4}$\,yrs. The uncertainty on these calculations has been calculated using Poisson statistics and is shown in Fig.\,\ref{fig:stat_lifetimes}, with the mean error being found to be $\sim$10\,per\,cent. As the Poisson errors rely on the number of clumps at each volume density (error = $\sqrt{N}$, where $N$ is the number of sources), the errors for the lifetimes are greatly increased towards higher volume densities due to the small number of clumps that possess these increase volume densities.

\subsection{``Straw man'' model comparison}

One of the main aims of this study is to investigate the \cite{Ellingsen2007b} ``straw man'' model using the physical properties presented in the ATLASGAL catalogue. This model is based on maser observations within regions of ongoing star formation and that, methanol masers (class I \& II) are associated with a very early stage of formation, followed by water masers. Hydroxyl masers are then seen to be coincident with ultra-compact \hii\ regions. In Sect.\,\ref{sect:evo_stage} we show that all of the maser associated clumps have similar \lmratio\ ratios, which we are using as our gauge of protostellar evolution.

These differences between the \lmratio\ ratios of the maser associated clumps can be used to investigate the ``straw man'' model. In Fig.\,\ref{fig:evo_stage_boxplot} we present a box plot of the distribution of \lmratio\ ratios for the maser associated clumps. In general, there is a good agreement between the results presented in Fig.\,\ref{fig:evo_stage_boxplot}, and the previous model presented in \cite{Breen2010}. However, a difference that can be noted is that water masers are seen towards regions of star formation before methanol masers and outlast hydroxyl masers, as predicted by the ``straw man'' model. While these findings show that the ``straw man'' model is fairly consistent with the \lmratio\ ratios of associated clumps,  it can be seen from Fig.\,\ref{fig:evo_stage_boxplot} that there is a considerably overlap between all of the maser species, which limits the power of this model as an evolutionary tracer for star formation.

We can also attempt to test the ``straw man'' model using the calculated statistical lifetimes for each maser phases. \cite{Breen2010} predicted that the lifetime of the 12.2\,GHz methanol maser to be between 1.5 $\times$ 10$^{4}$\,yrs and 2.7 $\times$ 10$^{4}$\,yrs. The statistical lifetime that we calculate for this maser does lie towards the centre of this range at $\sim$2$\times$10$^{4}$\,yrs, and so our results are in good agreement with \cite{Breen2010}. Along with the statistical lifetime for the 6.7\,GHz methanol maser predicted by \cite{VanDerWalt2005} (2.5$-$4.5$\times$10$^4$\,yrs) and the result found in \cite{Billington2019a} ($\sim$3.3$\times$10$^{4}$\,yrs) our results support the ``straw man'' model in terms of the methanol masers. The model presented by \cite{Breen2010} predicts that the hydroxyl maser has a lifetime of $\sim$20\,000\,yrs and the value found here is only a fifth of that prediction ($\sim$0.4$\times$10$^{4}$\,yrs). Finally, water masers were predicted to have relative lifetimes of $\sim$30\,000\,yrs, while we find a lifetime of approximately one half of this at only 16\,000\,yrs. One aspect that has not been included in this study is the approximate lifetime of ultra-compact \hii\ (UCHIIs) regions, and where the maser lifetimes lie in comparison to this. The ``straw man'' model predicts that UCHII regions begin to develop after the onset of water and 6.7\,GHz methanol masers, and exist throughout the lifetime of the 12.2\,GHz methanol masers and the hydroxyl masers. \cite{Kawamura1998} predicted the dynamical age of a W3(OH) to be approximately 2\,300 yrs, similar to our prediction of the hydroxyl maser lifetime but an order of magnitude less than the 12.2\,GHz statistical lifetime. Therefore, the ``straw man'' model and our results may be overestimating the lifetime of the 12.2\,GHz. However, it is currently unknown whether a UCHII region is required for the production of 12.2\,GHz methanol masers and for how long after the development of a UCHII region could methanol maser emission be sustained.

All of the lifetimes that are estimated in this study are only a lower limit to the true lifetimes for each maser species, as the lifetimes depend on the sensitivity and completeness of each of the maser surveys. As these lifetimes are a lower limit on the true maser lifetimes, we find that they are still in agreement with the \cite{Ellingsen2007b} model. While these measurements give a good indication of lifetimes for each maser species, we find it difficult to secure where these lifetimes lie in relation to one another as the \lmratio\ ratios show no significant trends. Overall we find a good agreement with the ``straw man'' model in terms of the statistical lifetimes of the maser species investigated, while having employed a different method to previous studies to calculate these lifetimes.

\section{Conclusions}
\label{sect:conclusions}

This study has investigated the correlations between the ATLASGAL catalogue and dense Galactic clumps that are associated with methanol, water and hydroxyl maser emission. We have used catalogues from the HOPS, THOR, SPLASH, MMB and ATLASGAL surveys, along with 12.2\,GHz MMB follow-up observations, to match maser emission to dense clumps located in the Galactic mid-plane ($|\ell|$ $<$ 60\degr\ and $|b|$ $<$ 1.5\degr). 

The association rates between maser emission and dust clumps for the 22.2\,GHz water and 12.2\,GHz methanol masers are found to be 56 and 82\,per\,cent respectively. The association rates for the hydroxyl masers were found to be: 3\,per\,cent (1612\,MHz), 60\,per\,cent (1665\,MHz), 42\,per\,cent (1667\,MHz), 49\,per\,cent  (1720\,MHz). Physical parameters for the maser associated clumps and the full ATLASGAL sample are taken from \cite{Urquhart2018} and \cite{Billington2019a}.

\begin{enumerate}

\item We find that the majority of methanol and water maser emission across the Galactic plane are associated with dense clumps, as identified by the ATLASGAL survey. Where a maser match has been found, they appear to have tightly correlated systematic velocities as those found for their counterpart clumps. The majority of masers ($\sim$90\,per\,cent) are also found to be tightly correlated with the peak of the dust emission ($<10$\,arcsec), where the highest densities are found. This implies that they are at least coincident if not directly associated with embedded star formation. 

\item It is just as common to find clumps coincident with multiple maser species ($\sim$45\,per\,cent) as those associated with only a single maser species ($\sim$55\,per\,cent). The communality of multiple species and/or transitions being found in a large fraction of clumps may be due to multiple evolutionary stages being present in each clump or that the various maser species require similar physical conditions. This is supported by the fact that all maser associated clumps, regardless of the corresponding maser species, have similar properties (we find no statistical differences between the clumps that are associated with different maser transitions).

\item Clumps associated with a maser are significantly more compact and dense than those that do not host any maser emission. However, we find no differences in radius and volume density between clumps which are associated with different maser species.

\item There is a similar density threshold required for the production of all species of maser emission as found in \cite{Billington2019a} (clump densities of greater than 10$^{4.1}$\,\cmthree), further justifying that volume density is an important factor for maser emission. Furthermore, the fraction of clumps with associated maser emission increases with volume density.

\item The \lmratio\ ratios of maser associated clumps, regardless of the associated maser emission, are shown to occupy the same distinct region of the parameter space, and so all types of maser emission can be seen to have similar turn-on and turn-off points in the evolutionary sequence. The ``straw man'' model \citep{Ellingsen2007b} predicts that the different maser species turn on and off at different times, however, large uncertainties associated with our results has limited any detailed comparison with the model but the \lmratio\ ratios are broadly consistent with previous findings.

\item We have constrained the physical properties required for maser emission and it is shown that masers only exist in clumps with volume densities above 10$^{4.1}$\,\cmthree, luminosities greater than $\sim$500\,\lsun\ and also require a minimum protostellar mass, estimated to be $\sim$6\,\msun. Maser species also have an approximate turn-on point in the evolutionary process of star formation ($\sim$10$^{0.2}$\,\lsun/\msun).

\item Statistical lifetimes are calculated for the water, hydroxyl and 12.2\,GHz methanol masers, and these lifetimes are found to be $\sim$1.6, 0.4 and 2.0$\times$10$^{4}$\,yrs respectively. We find that the lifetimes for the 6.7\,GHz (as found in \citealt{Billington2019a}) and 12.2\,GHz methanol masers is in good agreement with the values predicted by \cite{Breen2010}, whereas the statistical lifetimes determined for the water and hydroxyl masers are considerably shorter than those predicted, by one quarter and one half respectively. The lifetimes calculated for all masers are a lower limit on the true lifetimes, and so, our results support the ``straw man'' model \citep{Ellingsen2007b}.

\end{enumerate}

\section*{Acknowledgements}

S.\,J.\,Billington wishes to acknowledge an STFC (Science and Technology Facilities Council) PhD studentship for this work. H.\,Beuther acknowledges support from the European Research Council under the Horizon 2020 Framework Program via the ERC Consolidator Grant CSF-648505, and RSK via the ERC Advanced-Grant 339177 (STARLIGHT). H.\,Beuther further acknowledges support from the Deutsche Forschungsgemeinschaft in the Collaborative Research Center (SFB 881) "The Milky Way 74 System" (subproject B1). We have used the collaborative tool Overleaf available at: https://www.overleaf.com/.

\section*{Data Availability}

The data underlying this article will be shared on reasonable request to the corresponding author.



\bibliographystyle{mnras}
\bibliography{library,urquhart2016,manual}

\begin{thebibliography}{}
\makeatletter
\relax
\def\mn@urlcharsother{\let\do\@makeother \do\$\do\&\do\#\do\^\do\_\do\%\do\~}
\def\mn@doi{\begingroup\mn@urlcharsother \@ifnextchar [ {\mn@doi@}
  {\mn@doi@[]}}
\def\mn@doi@[#1]#2{\def\@tempa{#1}\ifx\@tempa\@empty \href
  {http://dx.doi.org/#2} {doi:#2}\else \href {http://dx.doi.org/#2} {#1}\fi
  \endgroup}
\def\mn@eprint#1#2{\mn@eprint@#1:#2::\@nil}
\def\mn@eprint@arXiv#1{\href {http://arxiv.org/abs/#1} {{\tt arXiv:#1}}}
\def\mn@eprint@dblp#1{\href {http://dblp.uni-trier.de/rec/bibtex/#1.xml}
  {dblp:#1}}
\def\mn@eprint@#1:#2:#3:#4\@nil{\def\@tempa {#1}\def\@tempb {#2}\def\@tempc
  {#3}\ifx \@tempc \@empty \let \@tempc \@tempb \let \@tempb \@tempa \fi \ifx
  \@tempb \@empty \def\@tempb {arXiv}\fi \@ifundefined
  {mn@eprint@\@tempb}{\@tempb:\@tempc}{\expandafter \expandafter \csname
  mn@eprint@\@tempb\endcsname \expandafter{\@tempc}}}

\bibitem[\protect\citeauthoryear{{Argon}, {Reid}  \& {Menten}}{{Argon}
  et~al.}{2000}]{Argon2000}
{Argon} A.~L.,  {Reid} M.~J.,   {Menten} K.~M.,  2000, \mn@doi [The
  Astrophysical Journal Supplement Series] {10.1086/313406}, \href
  {https://ui.adsabs.harvard.edu/abs/2000ApJS..129..159A} {129, 159}

\bibitem[\protect\citeauthoryear{{Batrla}, {Matthews}, {Menten}  \&
  {Walmsley}}{{Batrla} et~al.}{1987}]{Batrla1987}
{Batrla} W.,  {Matthews} H.~E.,  {Menten} K.~M.,   {Walmsley} C.~M.,  1987,
  \mn@doi [\nat] {10.1038/326049a0}, \href
  {https://ui.adsabs.harvard.edu/abs/1987Natur.326...49B} {326, 49}

\bibitem[\protect\citeauthoryear{Bertoldi \& McKee}{Bertoldi \&
  McKee}{1992}]{Bertoldi1992}
Bertoldi F.,  McKee C.~F.,  1992, \mn@doi [Astrophys. J.] {10.1086/171638},
  395, 140

\bibitem[\protect\citeauthoryear{Beuther, Walsh, Schilke, Sridharan, Menten  \&
  Wyrowski}{Beuther et~al.}{2002}]{Beuther2002}
Beuther H.,  Walsh A.,  Schilke P.,  Sridharan T.~K.,  Menten K.~M.,   Wyrowski
  F.,  2002, \mn@doi [Astron. Astrophys.] {10.1051/0004-6361:20020710}, 390,
  289

\bibitem[\protect\citeauthoryear{Beuther et~al.,}{Beuther
  et~al.}{2016}]{Beuther2016}
Beuther H.,  et~al., 2016, A{\&}A, 595, A32

\bibitem[\protect\citeauthoryear{Beuther et~al.,}{Beuther
  et~al.}{2019}]{Beuther2019}
Beuther H.,  et~al., 2019, \mn@doi [Astron. Astrophys.]
  {10.1051/0004-6361/201935936}, 628, A90

\bibitem[\protect\citeauthoryear{Bihr et~al.,}{Bihr et~al.}{2016}]{Bihr2016}
Bihr S.,  et~al., 2016, \mn@doi [A{\&}A] {10.1051/0004-6361/201527697}, 588,
  A97

\bibitem[\protect\citeauthoryear{Billington et~al.,}{Billington
  et~al.}{2019}]{Billington2019a}
Billington S.~J.,  et~al., 2019, \mn@doi [Mon. Not. R. Astron. Soc.]
  {10.1093/mnras/stz2691}, 2798, 2779

\bibitem[\protect\citeauthoryear{B{\l}aszkiewicz \& Kus}{B{\l}aszkiewicz \&
  Kus}{2004}]{Blaszkiewicz2004}
B{\l}aszkiewicz L.,  Kus A.~J.,  2004, \mn@doi [A{\&}A]
  {10.1051/0004-6361:20031451}, 413, 233

\bibitem[\protect\citeauthoryear{Breen \& Ellingsen}{Breen \&
  Ellingsen}{2011}]{Breen2011}
Breen S.~L.,  Ellingsen S.~P.,  2011, \mn@doi [MNRAS]
  {10.1111/j.1365-2966.2011.19020.x}, 416, 178

\bibitem[\protect\citeauthoryear{Breen, Ellingsen, Caswell  \& Lewis}{Breen
  et~al.}{2010}]{Breen2010}
Breen S.~L.,  Ellingsen S.~P.,  Caswell J.~L.,   Lewis B.~E.,  2010, \mn@doi
  [MNRAS] {10.1111/j.1365-2966.2009.15831.x}, 401, 2219

\bibitem[\protect\citeauthoryear{Breen, Ellingsen, Caswell, Green, Voronkov,
  Avison, Fuller  \& Quinn}{Breen et~al.}{2012a}]{Breen2012}
Breen S.~L.,  Ellingsen S.~P.,  Caswell J.~L.,  Green J.~A.,  Voronkov M.~A.,
  Avison A.,  Fuller G.~A.,   Quinn L.~J.,  2012a, \mn@doi [MNRAS]
  {10.1093/mnras/stw965}, 421, 1703

\bibitem[\protect\citeauthoryear{Breen, Ellingsen, Caswell, Green, Voronkov,
  Fuller, Quinn  \& Avison}{Breen et~al.}{2012b}]{Breen2012b}
Breen S.~L.,  Ellingsen S.~P.,  Caswell J.~L.,  Green J.~A.,  Voronkov M.~A.,
  Fuller G.~A.,  Quinn L.~J.,   Avison A.,  2012b, \mn@doi [MNRAS]
  {10.1111/j.1365-2966.2012.21759.x}, 426, 2189

\bibitem[\protect\citeauthoryear{Breen, Ellingsen, Contreras, Green, Caswell,
  Stevens, Dawson  \& Voronkov}{Breen et~al.}{2013}]{Breen2013}
Breen S.~L.,  Ellingsen S.~P.,  Contreras Y.,  Green J.~A.,  Caswell J.~L.,
  Stevens J.~B.,  Dawson J.~R.,   Voronkov M.~A.,  2013, \mn@doi [MNRAS]
  {10.1093/mnras/stt1315}, 435, 524

\bibitem[\protect\citeauthoryear{Breen et~al.,}{Breen
  et~al.}{2014}]{Breen2014a}
Breen S.~L.,  et~al., 2014, \mn@doi [MNRAS] {10.1093/mnras/stt2447}, 438, 3368

\bibitem[\protect\citeauthoryear{Breen et~al.,}{Breen et~al.}{2015}]{Breen2015}
Breen S.~L.,  et~al., 2015, \mn@doi [Mon. Not. R. Astron. Soc.]
  {10.1093/mnras/stv847}, 450, 4109

\bibitem[\protect\citeauthoryear{Breen, Ellingsen, Caswell, Green, Voronkov,
  Avison, Fuller  \& Quinn}{Breen et~al.}{2016}]{Breen2016a}
Breen S.~L.,  Ellingsen S.~P.,  Caswell J.~L.,  Green J.~A.,  Voronkov M.~A.,
  Avison A.,  Fuller G.~A.,   Quinn L.~J.,  2016, \mn@doi [MNRAS]
  {10.1093/mnras/stw965}, 459, 4066

\bibitem[\protect\citeauthoryear{Breen et~al.,}{Breen et~al.}{2018}]{Breen2018}
Breen S.~L.,  et~al., 2018, \mn@doi [MNRAS] {10.1093/mnras/stx3051}, 474, 3898

\bibitem[\protect\citeauthoryear{{Caswell}}{{Caswell}}{2004}]{Caswell2004}
{Caswell} J.~L.,  2004, \mn@doi [Monthly Notices of the Royal Astronomical
  Society] {10.1111/j.1365-2966.2004.07472.x}, \href
  {https://ui.adsabs.harvard.edu/abs/2004MNRAS.349...99C} {349, 99}

\bibitem[\protect\citeauthoryear{{Caswell}, {Vaile}, {Ellingsen}  \&
  {Norris}}{{Caswell} et~al.}{1995}]{Caswell1995b}
{Caswell} J.~L.,  {Vaile} R.~A.,  {Ellingsen} S.~P.,   {Norris} R.~P.,  1995,
  \mn@doi [\mnras] {10.1093/mnras/274.4.1126}, \href
  {https://ui.adsabs.harvard.edu/abs/1995MNRAS.274.1126C} {274, 1126}

\bibitem[\protect\citeauthoryear{Caswell et~al.,}{Caswell
  et~al.}{2010}]{Caswell2010}
Caswell J.~L.,  et~al., 2010, \mn@doi [MNRAS]
  {10.1111/j.1365-2966.2010.16339.x}, 404, 1029

\bibitem[\protect\citeauthoryear{Caswell et~al.,}{Caswell
  et~al.}{2011}]{Caswell2011a}
Caswell J.~L.,  et~al., 2011, \mn@doi [MNRAS]
  {10.1111/j.1365-2966.2011.19383.x}, 417, 1964

\bibitem[\protect\citeauthoryear{Caswell, Green  \& Phillips}{Caswell
  et~al.}{2013}]{Caswell2013}
Caswell J.~L.,  Green J.~A.,   Phillips C.~J.,  2013, \mn@doi [MNRAS]
  {10.1093/mnras/stt239}, 431, 1180

\bibitem[\protect\citeauthoryear{{Cheung}, {Rank}, {Townes}, {Thornton}  \&
  {Welch}}{{Cheung} et~al.}{1969}]{Cheung1969}
{Cheung} A.~C.,  {Rank} D.~M.,  {Townes} C.~H.,  {Thornton} D.~D.,   {Welch}
  W.~J.,  1969, \mn@doi [\nat] {10.1038/221626a0}, \href
  {https://ui.adsabs.harvard.edu/abs/1969Natur.221..626C} {221, 626}

\bibitem[\protect\citeauthoryear{Churchwell et~al.,}{Churchwell
  et~al.}{2009}]{Churchwell2009}
Churchwell E.,  et~al., 2009, \mn@doi [Publ. Astron. Soc. Pacific]
  {10.1086/597811}, 121, 213

\bibitem[\protect\citeauthoryear{Claussen, Wilking, Benson, Wootten, Myers  \&
  Terebey}{Claussen et~al.}{1996}]{Claussen1996}
Claussen M.~J.,  Wilking B.~A.,  Benson P.~J.,  Wootten A.,  Myers P.~C.,
  Terebey S.,  1996, \mn@doi [Astrophys. J. Suppl. v.106] {10.1086/192330},
  106, 111

\bibitem[\protect\citeauthoryear{{Claussen}, {Goss}, {Frail}  \&
  {Desai}}{{Claussen} et~al.}{1999}]{Claussen1999}
{Claussen} M.~J.,  {Goss} W.~M.,  {Frail} D.~A.,   {Desai} K.,  1999, \mn@doi
  [The Astrophysical Journal] {10.1086/307641}, \href
  {https://ui.adsabs.harvard.edu/abs/1999ApJ...522..349C} {522, 349}

\bibitem[\protect\citeauthoryear{Codella, Lorenzani, Gallego, Cesaroni  \&
  Moscadelli}{Codella et~al.}{2004}]{Codella2004}
Codella C.,  Lorenzani A.,  Gallego A.~T.,  Cesaroni R.,   Moscadelli L.,
  2004, \mn@doi [A{\&}A] {10.1051/0004-6361:20035608}, 417, 615

\bibitem[\protect\citeauthoryear{Contreras et~al.,}{Contreras
  et~al.}{2013}]{Contreras2013}
Contreras Y.,  et~al., 2013, \mn@doi [A{\&}A] {10.1051/0004-6361/201220155},
  549, A45

\bibitem[\protect\citeauthoryear{Cragg, Sobolev, Ellingsen, Caswell, Godfrey,
  Salii  \& Dodson}{Cragg et~al.}{2001}]{Cragg2001}
Cragg D.~M.,  Sobolev A.~M.,  Ellingsen S.~P.,  Caswell J.~L.,  Godfrey P.~D.,
  Salii S.~V.,   Dodson R.~G.,  2001, \mn@doi [Mon. Not. R. Astron. Soc.]
  {10.1046/j.1365-8711.2001.04294.x}, 323, 939

\bibitem[\protect\citeauthoryear{Cragg, Sobolev  \& Godfrey}{Cragg
  et~al.}{2002}]{Cragg2002a}
Cragg D.~M.,  Sobolev A.~M.,   Godfrey P.~D.,  2002, \mn@doi [Mon. Not. R.
  Astron. Soc.] {10.1046/j.1365-8711.2002.05226.x}, 331, 521

\bibitem[\protect\citeauthoryear{Csengeri et~al.,}{Csengeri
  et~al.}{2014}]{Csengeri2014}
Csengeri T.,  et~al., 2014, \mn@doi [A{\&}A] {10.1051/0004-6361/201322434},
  565, A75

\bibitem[\protect\citeauthoryear{Dawson et~al.,}{Dawson
  et~al.}{2014}]{Dawson2014}
Dawson J.~R.,  et~al., 2014, \mn@doi [Mon. Not. R. Astron. Soc.]
  {10.1093/mnras/stu032}, 439, 1596

\bibitem[\protect\citeauthoryear{Eden et~al.,}{Eden et~al.}{2017}]{Eden2017}
Eden D.~J.,  et~al., 2017, \mn@doi [MNRAS] {10.1093/mnras/stx874}, 469, 2163

\bibitem[\protect\citeauthoryear{Elitzur}{Elitzur}{1992}]{Elitzur1992}
Elitzur M.,  1992, \mn@doi [ARA] {10.1126/science.1101353}, 30, 75

\bibitem[\protect\citeauthoryear{{Elitzur}, {Goldreich}  \&
  {Scoville}}{{Elitzur} et~al.}{1976}]{Elitzur1976}
{Elitzur} M.,  {Goldreich} P.,   {Scoville} N.,  1976, \mn@doi [The
  Astrophysical Journal] {10.1086/154289}, \href
  {https://ui.adsabs.harvard.edu/abs/1976ApJ...205..384E} {205, 384}

\bibitem[\protect\citeauthoryear{Elitzur, Hollenbach  \& Mckee}{Elitzur
  et~al.}{1989}]{Elitzur1989}
Elitzur M.,  Hollenbach D.~J.,   Mckee C.~F.,  1989, ApJ, 346, 983

\bibitem[\protect\citeauthoryear{Ellingsen, Voronkov, Cragg  \&
  Sobolev}{Ellingsen et~al.}{2007}]{Ellingsen2007b}
Ellingsen S.~P.,  Voronkov M.~A.,  Cragg D.~M.,   Sobolev A.~M.,  2007, \mn@doi
  [Proc. Int. Astron. Union] {10.1017/S1743921307012999}, 242, 213

\bibitem[\protect\citeauthoryear{{Forster} \& {Caswell}}{{Forster} \&
  {Caswell}}{1989}]{Forster1989}
{Forster} J.~R.,  {Caswell} J.~L.,  1989, \aap, \href
  {https://ui.adsabs.harvard.edu/abs/1989A&A...213..339F} {213, 339}

\bibitem[\protect\citeauthoryear{Goedhart, Gaylard  \& {Van Der Walt}}{Goedhart
  et~al.}{2005}]{Goedhart2005}
Goedhart S.,  Gaylard M.~J.,   {Van Der Walt} D.~J.,  2005, \mn@doi [Astrophys.
  Space Sci.] {10.1007/s10509-005-3688-8}, 295, 197

\bibitem[\protect\citeauthoryear{{Green}}{{Green}}{2009}]{Green2009}
{Green} J.~A. a.~a.,  2009, \mn@doi [\mnras]
  {10.1111/j.1365-2966.2008.14091.x}, \href
  {http://cdsads.u-strasbg.fr/abs/2009MNRAS.392..783G} {392, 783}

\bibitem[\protect\citeauthoryear{Green et~al.,}{Green
  et~al.}{2009}]{Green2009a}
Green J.~A.,  et~al., 2009, \mn@doi [MNRAS] {10.1111/j.1365-2966.2008.14091.x},
  392, 783

\bibitem[\protect\citeauthoryear{Green et~al.,}{Green et~al.}{2010}]{Green2010}
Green J.~A.,  et~al., 2010, \mn@doi [MNRAS] {10.1111/j.1365-2966.2010.17376.x},
  409, 913

\bibitem[\protect\citeauthoryear{Green et~al.,}{Green et~al.}{2012}]{Green2012}
Green J.~A.,  et~al., 2012, \mn@doi [MNRAS] {10.1111/j.1365-2966.2011.20229.x},
  420, 3108

\bibitem[\protect\citeauthoryear{{Guzm{\'a}n}, {Garay}, {Brooks}  \&
  {Voronkov}}{{Guzm{\'a}n} et~al.}{2012}]{Guzman2012}
{Guzm{\'a}n} A.~E.,  {Garay} G.,  {Brooks} K.~J.,   {Voronkov} M.~A.,  2012,
  \mn@doi [\apj] {10.1088/0004-637X/753/1/51}, \href
  {https://ui.adsabs.harvard.edu/abs/2012ApJ...753...51G} {753, 51}

\bibitem[\protect\citeauthoryear{{Henning}, {Lapinov}, {Schreyer}, {Stecklum}
  \& {Zinchenko}}{{Henning} et~al.}{2000}]{Henning2000}
{Henning} T.,  {Lapinov} A.,  {Schreyer} K.,  {Stecklum} B.,   {Zinchenko} I.,
  2000, \aap, \href {https://ui.adsabs.harvard.edu/abs/2000A&A...364..613H}
  {364, 613}

\bibitem[\protect\citeauthoryear{Hill, Burton, Minier, Thompson, Walsh,
  Hunt-Cunningham  \& Garay}{Hill et~al.}{2005}]{Hill2005}
Hill T.,  Burton M.~G.,  Minier V.,  Thompson M.~A.,  Walsh A.~J.,
  Hunt-Cunningham M.,   Garay G.,  2005, \mn@doi [MNRAS]
  {10.1111/j.1365-2966.2005.09347.x}, 363, 405

\bibitem[\protect\citeauthoryear{{Jones} et~al.,}{{Jones}
  et~al.}{2020}]{Jones2020}
{Jones} B.~M.,  et~al., 2020, \mn@doi [\mnras] {10.1093/mnras/staa233}, \href
  {https://ui.adsabs.harvard.edu/abs/2020MNRAS.493.2015J} {493, 2015}

\bibitem[\protect\citeauthoryear{Kauffmann, Pillai  \& Goldsmith}{Kauffmann
  et~al.}{2013}]{Kauffmann2013a}
Kauffmann J.,  Pillai T.,   Goldsmith P.~F.,  2013, \mn@doi [Astrophys. J.]
  {10.1088/0004-637X/779/2/185}, 779

\bibitem[\protect\citeauthoryear{{Kawamura} \& {Masson}}{{Kawamura} \&
  {Masson}}{1998}]{Kawamura1998}
{Kawamura} J.~H.,  {Masson} C.~R.,  1998, \mn@doi [\apj] {10.1086/306472},
  \href {https://ui.adsabs.harvard.edu/abs/1998ApJ...509..270K} {509, 270}

\bibitem[\protect\citeauthoryear{K{\"{o}}nig et~al.,}{K{\"{o}}nig
  et~al.}{2017}]{Konig2017}
K{\"{o}}nig C.,  et~al., 2017, \mn@doi [A{\&}A] {10.1051/0004-6361/201526841},
  599, 1

\bibitem[\protect\citeauthoryear{Kuiper}{Kuiper}{1938}]{Kuiper1938}
Kuiper G.~P.,  1938, ApJ, 88, 472

\bibitem[\protect\citeauthoryear{Menten}{Menten}{1991}]{Menten1991}
Menten K.~M.,  1991, \mn@doi [ApJ] {10.1086/186177}, 380, L75

\bibitem[\protect\citeauthoryear{{Menten}}{{Menten}}{1997}]{Menten1997}
{Menten} K.~M.,  1997, in {van Dishoeck} E.~F.,  ed.,  IAU Symposium Vol. 178,
  IAU Symposium. pp 163--172, \mn@doi{10.1017/S0074180900009323}

\bibitem[\protect\citeauthoryear{{Menten}, {Reid}, {Pratap}, {Moran}  \&
  {Wilson}}{{Menten} et~al.}{1992}]{Menten1992}
{Menten} K.~M.,  {Reid} M.~J.,  {Pratap} P.,  {Moran} J.~M.,   {Wilson} T.~L.,
  1992, \mn@doi [\apjl] {10.1086/186665}, \href
  {https://ui.adsabs.harvard.edu/abs/1992ApJ...401L..39M} {401, L39}

\bibitem[\protect\citeauthoryear{Minier}{Minier}{2000}]{Minier2000}
Minier V.,  2000, Astron. Astrophys., 362, 1093

\bibitem[\protect\citeauthoryear{Minier, Ellingsen, Norris  \& Booth}{Minier
  et~al.}{2003}]{Minier2003}
Minier V.,  Ellingsen S.~P.,  Norris R.~P.,   Booth R.~S.,  2003, \mn@doi
  [A{\&}A] {10.1051/0004-6361:20030465}, 403, 1095

\bibitem[\protect\citeauthoryear{Molinari, Pezzuto, Cesaroni, Brand, Faustini
  \& Testi}{Molinari et~al.}{2008}]{Molinari2008}
Molinari S.,  Pezzuto S.,  Cesaroni R.,  Brand J.,  Faustini F.,   Testi L.,
  2008, \mn@doi [A{\&}A] {10.1051/0004-6361:20078661}, 481, 345

\bibitem[\protect\citeauthoryear{Molinari et~al.,}{Molinari
  et~al.}{2010}]{Molinari2010}
Molinari S.,  et~al., 2010, \mn@doi [Publ. Astron. Soc. Pacific]
  {10.1086/651314}, 122, 314

\bibitem[\protect\citeauthoryear{Molinari et~al.,}{Molinari
  et~al.}{2019}]{Molinari2019}
Molinari S.,  et~al., 2019, \mn@doi [Mon. Not. R. Astron. Soc.]
  {10.1093/mnras/stz900}, 486, 4508

\bibitem[\protect\citeauthoryear{Moore et~al.,}{Moore et~al.}{2015}]{Moore2015}
Moore T.~J.,  et~al., 2015, \mn@doi [MNRAS] {10.1093/mnras/stv1833}, 453, 4264

\bibitem[\protect\citeauthoryear{{Motte}, {Schilke}  \& {Lis}}{{Motte}
  et~al.}{2002}]{Motte2002}
{Motte} F.,  {Schilke} P.,   {Lis} D.~C.,  2002, {Massive Star Formation in the
  Galactic Mini-Starburst W43}.
p.~393

\bibitem[\protect\citeauthoryear{Norris et~al.,}{Norris
  et~al.}{1998}]{Norris1998}
Norris R.~P.,  et~al., 1998, \mn@doi [ApJ] {10.1086/306373}, 508, 275

\bibitem[\protect\citeauthoryear{Phillips, Norris, Ellingsen  \&
  McCulloch}{Phillips et~al.}{1998}]{Phillips1998}
Phillips C.~J.,  Norris R.~P.,  Ellingsen S.~P.,   McCulloch P.~M.,  1998,
  \mn@doi [MNRAS] {10.1046/j.1365-8711.1998.01979.x}, 300, 1131

\bibitem[\protect\citeauthoryear{{Qiao}, {Li}, {Shen}, {Chen}  \&
  {Zheng}}{{Qiao} et~al.}{2014}]{Qiao2014}
{Qiao} H.,  {Li} J.,  {Shen} Z.,  {Chen} X.,   {Zheng} X.,  2014, \mn@doi
  [Monthly Notices of the Royal Astronomical Society] {10.1093/mnras/stu776},
  \href {https://ui.adsabs.harvard.edu/abs/2014MNRAS.441.3137Q} {441, 3137}

\bibitem[\protect\citeauthoryear{Qiao et~al.,}{Qiao et~al.}{2016}]{Qiao2016}
Qiao H.-H.,  et~al., 2016, \mn@doi [Astrophys. J. Suppl. Ser.]
  {10.3847/1538-4365/227/2/26}, 227, 26

\bibitem[\protect\citeauthoryear{Qiao et~al.,}{Qiao et~al.}{2018}]{Qiao2018}
Qiao H.-H.,  et~al., 2018, \mn@doi [Astrophys. J. Suppl. Ser.]
  {10.3847/1538-4365/aae580}, 239, 15

\bibitem[\protect\citeauthoryear{Reid, Dame, Menten  \& Brunthaler}{Reid
  et~al.}{2016}]{Reid2016}
Reid M.~J.,  Dame T.~M.,  Menten K.~M.,   Brunthaler A.,  2016, \mn@doi
  [Astrophys. J.] {10.3847/0004-637X/823/2/77}, 823, 1

\bibitem[\protect\citeauthoryear{Rosolowsky et~al.,}{Rosolowsky
  et~al.}{2010}]{Rosolowsky2010}
Rosolowsky E.,  et~al., 2010, \mn@doi [Astrophys. Journal, Suppl. Ser.]
  {10.1088/0067-0049/188/1/123}, 188, 123

\bibitem[\protect\citeauthoryear{Schuller et~al.,}{Schuller
  et~al.}{2009}]{Schuller2009}
Schuller F.,  et~al., 2009, \mn@doi [A{\&}A] {10.1051/0004-6361/200811568},
  504, 415

\bibitem[\protect\citeauthoryear{Siringo et~al.,}{Siringo
  et~al.}{2009}]{Siringo2009}
Siringo G.,  et~al., 2009, \mn@doi [A{\&}A] {10.1117/12.787981}, 497, 945

\bibitem[\protect\citeauthoryear{Sobolev, Cragg  \& Godfrey}{Sobolev
  et~al.}{1997}]{Sobolev1997}
Sobolev A.~M.,  Cragg D.~M.,   Godfrey P.~D.,  1997, A{\&}A, 324, 211

\bibitem[\protect\citeauthoryear{Szymczak, Kus, Hrynek, Kepa  \&
  Pazderski}{Szymczak et~al.}{2002}]{Szymczak2002}
Szymczak M.,  Kus A.~J.,  Hrynek G.,  Kepa A.,   Pazderski E.,  2002, 392, 277

\bibitem[\protect\citeauthoryear{{Szymczak}, {Pillai}  \& {Menten}}{{Szymczak}
  et~al.}{2005}]{Szymczak2005}
{Szymczak} M.,  {Pillai} T.,   {Menten} K.~M.,  2005, \mn@doi [\aap]
  {10.1051/0004-6361:20042437}, \href
  {https://ui.adsabs.harvard.edu/abs/2005A&A...434..613S} {434, 613}

\bibitem[\protect\citeauthoryear{Thomasson}{Thomasson}{1986}]{Thomasson1986}
Thomasson P.,  1986, Q. J. R. Astron. Soc., 27, 413

\bibitem[\protect\citeauthoryear{Titmarsh, Ellingsen, Breen, Caswell  \&
  Voronkov}{Titmarsh et~al.}{2014}]{Titmarsh2014}
Titmarsh A.~M.,  Ellingsen S.~P.,  Breen S.~L.,  Caswell J.~L.,   Voronkov
  M.~A.,  2014, \mn@doi [MNRAS] {10.1093/mnras/stw636}, 443, 2923

\bibitem[\protect\citeauthoryear{Titmarsh, Ellingsen, Breen, Caswell  \&
  Voronkov}{Titmarsh et~al.}{2016}]{Titmarsh2016}
Titmarsh A.~M.,  Ellingsen S.~P.,  Breen S.~L.,  Caswell J.~L.,   Voronkov
  M.~A.,  2016, \mn@doi [MNRAS] {10.1093/mnras/stw636}, 459, 157

\bibitem[\protect\citeauthoryear{Urquhart et~al.,}{Urquhart
  et~al.}{2011}]{Urquhart2011a}
Urquhart J.~S.,  et~al., 2011, \mn@doi [MNRAS]
  {10.1111/j.1365-2966.2011.19594.x}, 418, 1689

\bibitem[\protect\citeauthoryear{Urquhart et~al.,}{Urquhart
  et~al.}{2013}]{Urquhart2013}
Urquhart J.~S.,  et~al., 2013, \mn@doi [MNRAS] {10.1093/mnras/stt287}, 431,
  1752

\bibitem[\protect\citeauthoryear{Urquhart et~al.,}{Urquhart
  et~al.}{2014a}]{Urquhart2014b}
Urquhart J.~S.,  et~al., 2014a, \mn@doi [MNRAS] {10.1093/mnras/stu1207}, 443,
  1555

\bibitem[\protect\citeauthoryear{Urquhart et~al.,}{Urquhart
  et~al.}{2014b}]{Urquhart2014a}
Urquhart J.~S.,  et~al., 2014b, \mn@doi [A{\&}A] {10.1051/0004-6361/201424126},
  568, A41

\bibitem[\protect\citeauthoryear{Urquhart et~al.,}{Urquhart
  et~al.}{2015}]{Urquhart2015}
Urquhart J.~S.,  et~al., 2015, \mn@doi [MNRAS] {10.1093/mnras/stu2300}, 446,
  3461

\bibitem[\protect\citeauthoryear{Urquhart et~al.,}{Urquhart
  et~al.}{2018}]{Urquhart2018}
Urquhart J.~S.,  et~al., 2018, \mn@doi [MNRAS] {10.1093/mnras/stx2258}, 473,
  1059

\bibitem[\protect\citeauthoryear{{Van Der Walt}}{{Van Der
  Walt}}{2005}]{VanDerWalt2005}
{Van Der Walt} J.,  2005, \mn@doi [MNRAS] {10.1111/j.1365-2966.2005.09026.x},
  360, 153

\bibitem[\protect\citeauthoryear{Walsh, Burton, Hyland  \& Robinson}{Walsh
  et~al.}{1998}]{Walsh1998}
Walsh A.~J.,  Burton M.~G.,  Hyland A.~R.,   Robinson G.,  1998, \mn@doi
  [MNRAS] {10.1111/j.1365-8711.1998.02014.x}, 301, 640

\bibitem[\protect\citeauthoryear{Walsh et~al.,}{Walsh et~al.}{2011}]{Walsh2011}
Walsh A.~J.,  et~al., 2011, Sci. Technol., 000

\bibitem[\protect\citeauthoryear{Walsh, Purcell, Longmore, Breen, Green,
  Harvey-Smith, Jordan  \& Macpherson}{Walsh et~al.}{2014}]{Walsh2014}
Walsh A.~J.,  Purcell C.~R.,  Longmore S.~N.,  Breen S.~L.,  Green J.~A.,
  Harvey-Smith L.,  Jordan C.~H.,   Macpherson C.,  2014, \mn@doi [MNRAS]
  {10.1093/mnras/stu989}, 442, 2240

\bibitem[\protect\citeauthoryear{Walsh et~al.,}{Walsh et~al.}{2016}]{Walsh2016}
Walsh A.~J.,  et~al., 2016, \mn@doi [MNRAS] {10.1093/mnras/stv2446}, 455, 3494

\bibitem[\protect\citeauthoryear{{Wang} et~al.,}{{Wang}
  et~al.}{2018}]{Wang2018}
{Wang} Y.,  et~al., 2018, \mn@doi [\aap] {10.1051/0004-6361/201833642}, \href
  {https://ui.adsabs.harvard.edu/abs/2018A&A...619A.124W} {619, A124}

\bibitem[\protect\citeauthoryear{{Weinreb}, {Barrett}, {Meeks}  \&
  {Henry}}{{Weinreb} et~al.}{1963}]{Weinreb1963}
{Weinreb} S.,  {Barrett} A.~H.,  {Meeks} M.~L.,   {Henry} J.~C.,  1963, \mn@doi
  [\nat] {10.1038/200829a0}, \href
  {https://ui.adsabs.harvard.edu/abs/1963Natur.200..829W} {200, 829}

\bibitem[\protect\citeauthoryear{Wienen, Wyrowski, Schuller, Menten, Walmsley,
  Bronfman  \& Motte}{Wienen et~al.}{2012}]{Wienen2012}
Wienen M.,  Wyrowski F.,  Schuller F.,  Menten K.~M.,  Walmsley C.~M.,
  Bronfman L.,   Motte F.,  2012, \mn@doi [A{\&}A]
  {10.1051/0004-6361/201118107}, 544, A146

\bibitem[\protect\citeauthoryear{Wienen, Wyrowski, Menten, Urquhart, Walmsley,
  Csengeri, Koribalski  \& Schuller}{Wienen et~al.}{2018}]{Wienen2018}
Wienen M.,  Wyrowski F.,  Menten K.~M.,  Urquhart J.~S.,  Walmsley C.~M.,
  Csengeri T.,  Koribalski B.~S.,   Schuller F.,  2018, \mn@doi [A{\&}A]
  {10.1051/0004-6361/201526384}, 609, A125

\bibitem[\protect\citeauthoryear{{Wilson} \& {Barrett}}{{Wilson} \&
  {Barrett}}{1968}]{Wilson1968}
{Wilson} W.~J.,  {Barrett} A.~H.,  1968, The Astronomical Journal Supplement,
  \href {https://ui.adsabs.harvard.edu/abs/1968AJS....73R.209W} {73, 209}

\bibitem[\protect\citeauthoryear{Xu, Li, Haschisuka, Pandian, Menten  \&
  Henkel}{Xu et~al.}{2008}]{Xu2008}
Xu Y.,  Li J.~J.,  Haschisuka K.,  Pandian J.~D.,  Menten K.~M.,   Henkel C.,
  2008, A{\&}A, 485, 729

\bibitem[\protect\citeauthoryear{Yang et~al.,}{Yang et~al.}{2017}]{Yang2017}
Yang K.,  et~al., 2017, \mn@doi [ApJ] {10.3847/1538-4357/aa8668}, 846, 160

\bibitem[\protect\citeauthoryear{Yang et~al.,}{Yang et~al.}{2019}]{Yang2019}
Yang K.,  et~al., 2019, \mn@doi [Astrophys. J. Suppl. Ser.]
  {10.3847/1538-4357/aa8668}, 241, 18

\makeatother
\end{thebibliography}




\bsp	
\label{lastpage}

\end{document}